\documentclass[final,3p,times, authoryear]{elsarticle}
\usepackage{graphicx}
\usepackage{subcaption}
\usepackage[export]{adjustbox}

\usepackage{amssymb}
\usepackage{amsthm}
\usepackage{amsmath}

\biboptions{sort&compress}
\usepackage[dvipsnames]{xcolor}
\usepackage{fullpage}
\usepackage{times}
\usepackage{fancyhdr,graphicx,amsmath,amssymb}
\usepackage[linesnumbered,ruled,vlined]{algorithm2e}
\usepackage{siunitx}
\usepackage{mathtools}
\usepackage{bbold}
\usepackage{pgfplots}
\pgfplotsset{width=(\columnwidth - 8pt),compat=1.9}
\usepackage{textcomp}
\usepackage{booktabs}
\usepackage{multirow}
\usepackage{caption}
\usepackage{tabularx}
\usepackage{placeins}
\usepackage{threeparttable}
\usepackage{soul}
\usepackage{float}
\restylefloat{table}

\newcolumntype{L}[1]{>{\raggedright\let\newline\\\arraybackslash\hspace{0pt}}m{#1}}
\newcolumntype{C}[1]{>{\centering\let\newline\\\arraybackslash\hspace{0pt}}m{#1}}
\newcolumntype{R}[1]{>{\raggedleft\let\newline\\\arraybackslash\hspace{0pt}}m{#1}}

\newcommand\var{5pt}

\usepackage{gensymb}
\usepackage{appendix}
\usepackage{multicol}
\usepackage[inline]{enumitem}
\usepackage{longtable}

\makeatletter
\def\ps@pprintTitle{%
  \let\@oddhead\@empty
  \let\@evenhead\@empty
  \let\@oddfoot\@empty
  \let\@evenfoot\@oddfoot
}
\makeatother

\begin{document}
\begin{frontmatter}

\title{Towards Online Monitoring and Data-driven Control: A Study of Segmentation Algorithms for Laser Powder Bed Fusion Processes}



\author[labelME]{Alexander Nettekoven}
\ead{nettekoven@utexas.edu}

\author[labelME]{Scott Fish}
\ead{scott.fish@utexas.edu}

\author[labelME]{Joseph Beaman}
\ead{jbeaman@me.utexas.edu}

\author[labelASE]{Ufuk Topcu}
\ead{utopcu@utexas.edu}

\address[labelME]{Department of Mechanical Engineering, The University of Texas at Austin, 204 E. Dean Keeton Street, Stop C2200, Austin, Texas 78712, USA}
\address[labelASE]{Department of Aerospace Engineering and Engineering Mechanics, The University of Texas at Austin, 2617 Wichita Street, C0600, Austin, Texas 78712, USA}

\begin{abstract}
An increasing number of laser powder bed fusion machines use off-axis infrared cameras to improve online monitoring and data-driven control capabilities. However, there is still a severe lack of algorithmic solutions to properly process the infrared images from these cameras that has led to several key limitations: a lack of online monitoring capabilities for the laser tracks, insufficient pre-processing of the infrared images for data-driven methods, and large memory requirements for storing the infrared images. To address these limitations, we study over 30 segmentation algorithms that segment each infrared image into a foreground and background. By evaluating each algorithm based on its segmentation accuracy, computational speed, and spatter detection characteristics, we identify promising algorithmic solutions. The identified algorithms can be readily applied to the laser powder bed fusion machines to address each of the above limitations and thus, significantly improve process control.
\end{abstract}


\end{frontmatter}


\section{Introduction} \label{section_1}

Numerous applications, such as aircraft and medical devices, increasingly rely on 3D parts produced by laser powder bed fusion (LPBF)~\citep{2_slm_advantage,3_SLS_SLM_application,4_SLS_SLM_appl_2,5_am_review_market}. Their potential of producing lightweight 3D parts with complex geometries creates a significant advantage for LPBF processes over traditional manufacturing processes~\citep{2_slm_advantage,1_trad_manufact}. However, limited process control leaves LPBF machines still susceptible to process disturbances that often lead to poor part quality~\citep{5_am_review_market,1_trad_manufact, 6_am_review_control, data_storage_problem}. Building high-quality parts and doing so repeatably, especially for safety-critical applications, remain a key challenge for LPBF processes today~\citep{4_SLS_SLM_appl_2, 5_am_review_market}.

\begin{figure}[h]
\centering
\begin{subfigure}{.35\textwidth}
  \centering
  \includegraphics[height=0.85\linewidth, trim={2.1cm 2cm 4.3cm 1.5cm},clip]{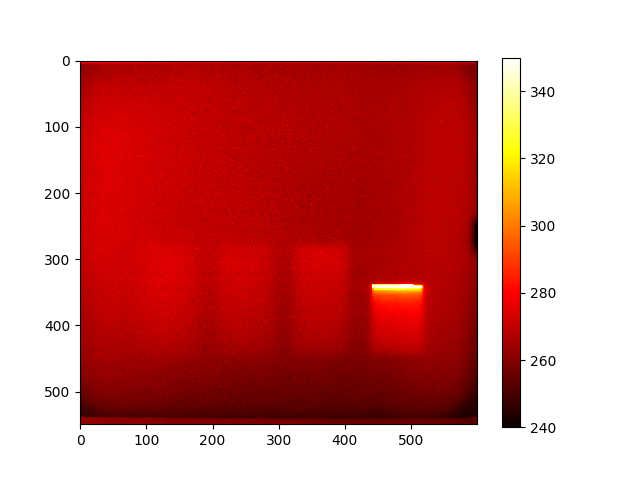}
  \caption{SLS off-axis infrared image.}
  \label{fig:full_view_1}
\end{subfigure}%
\begin{subfigure}{.35\textwidth}
  \centering
  \includegraphics[height=0.85\linewidth, trim={0cm 0cm 0cm 0cm},clip]{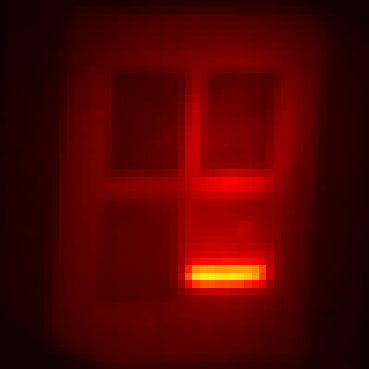}
  \caption{SLM off-axis infrared image.}
  \label{fig:full_view_2}
\end{subfigure}
\caption{Sample infrared images of build surfaces from two different builds used for this study, one from a selective laser sintering (SLS) machine and one from a selective laser melting (SLM) machine. The laser in both builds scanned four rectangular cross-sections for each layer.}
\label{fig:full_view_off_axis}
\end{figure}

\begin{figure}
\centering
  \includegraphics[width=0.7\columnwidth, trim={2cm 4cm 2cm 4cm},clip]{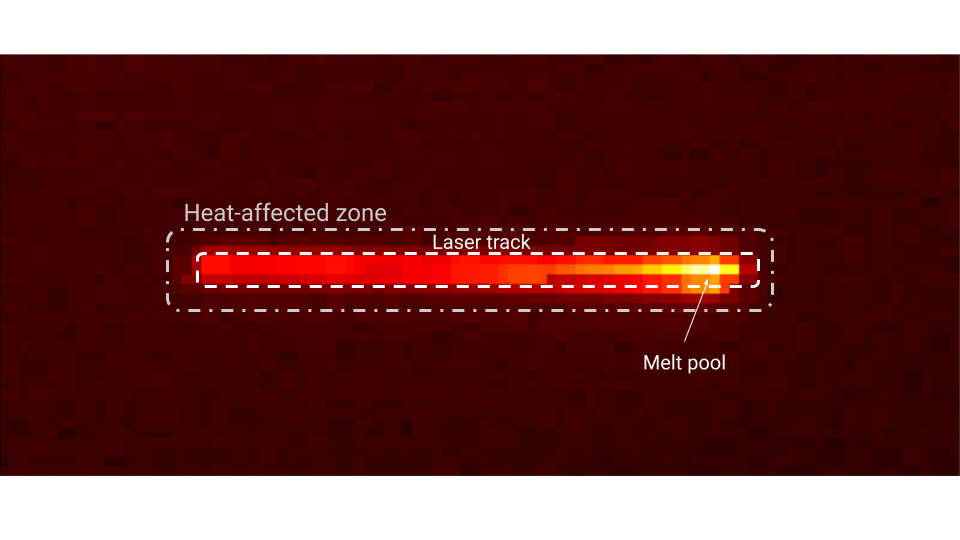}
\caption{A single laser track scanned by a laser. The approximate locations of several region of interests are marked. The cropped image is taken from the same SLS build as Figure \ref{fig:full_view_1}.}
\label{fig:first_scan_line_ROI}
\end{figure}

To improve process control, extensive research efforts are going into using in situ measurements from off-axis infrared cameras, which continuously record the entire build surface at once~\citep{SLM_review_insitu_monitoring, slm_NIST_cited_this_successor}. The infrared images produced by these cameras usually consist of \mbox{2D} arrays of temperature intensity values, where each value represents the measured temperature of the corresponding build surface location. Figure~\ref{fig:full_view_off_axis} shows two exemplary infrared images for the two type of LPBF processes, selective laser sintering (SLS) and selective laser melting (SLM).  In contrast to in situ measurements from other infrared sensors, infrared images from off-axis infrared cameras have the advantage of providing information about multiple regions of interest on the build surface at once. Some of the previously studied regions of interest include the melt pool, the plume, and the laser scans (see Figure~\ref{fig:first_scan_line_ROI})~\citep{SLM_review_insitu_monitoring,insitu_monitoring_SLM_zinc_powder,Zinc_successor_paper,slm_NIST_cited_this,slm_NIST_cited_this_successor, study_of_aluminum, optimized_hatch_space, effect_process_params_on_track, effect_process_params_on_track_2, effect_on_single_track_stability_zones, effect_of_hatch_length, effect_hatch_angle_on_parts}. Measuring the regions of interest can provide crucial information for process control~\citep{SLM_review_insitu_monitoring,insitu_monitoring_SLM_zinc_powder, Zinc_successor_paper, slm_NIST_cited_this,slm_NIST_cited_this_successor, effect_of_hatch_length, study_of_aluminum}. However, most LPBF processes still lack appropriate algorithms that can automatically extract and use this information from the infrared images of the build surface effectively~\citep{slm_review, SLM_ML_review, SLM_review_insitu_monitoring}.

In an effort to reduce the shortcoming of algorithmic solutions for these infrared images, recent studies have focused on developing different online monitoring and data-driven algorithms.~\citet{insitu_monitoring_SLM_zinc_powder} applied common computer vision techniques to identify the plume in each image and use the area and the mean intensity of the extracted plume to monitor the process health. In a follow-up study,~\citet{Zinc_successor_paper} improved upon this concept by developing statistical descriptors of the plume.~\citet{segmentation_2} developed an image segmentation algorithm that extracts the melt pool, plume, and spatters from the infrared images. The extracted regions were then monitored to study the effect of different process parameters.~\citet{segmentation_4} performed a similar process parameter study by extracting the plume and spatter from the infrared images with a popular image segmentation method called Otsu's method~\citep{Otsu}. Some other approaches used the whole image as input for data-driven methods;~\citet{DL_CNN} trained a convolutional neural network to detect splatter and deformation defects.~\citet{Zhe} applied a data-driven method that provided human-interpretable information about the cause of strength differences in test specimen.

Though these studies provide promising results for improving process control, there still exist severe limitations that have not been fully investigated or have not been addressed at all. 

\textbf{Monitoring limitation.} Recent developments of online monitoring and data-driven algorithms have mostly focused on the melt pool, the plume, the spatters, or the whole build surface. So far, the laser tracks, which are the result of the laser beam continuously traveling along the scan path on the build surface, have not been subject to any significant algorithmic developments. Extracting the laser tracks from the infrared images is difficult because the laser tracks' temperature distribution lies in between the temperature distribution of the melt pool and the temperature distribution of the rest of the powder bed, as indicated by Figure~\ref{fig:first_scan_line_ROI}. Monitoring a single or multiple laser tracks can provide crucial process information, such as geometrical accuracy of the scanned-cross section, deviation of process parameters, and consistency of each laser track~\citep{study_of_aluminum, effect_on_single_track_stability_zones, effect_process_params_on_track, effect_process_params_on_track_2}. Identifying the laser tracks in the infrared images may also enable monitoring of the heat-affected zone, which encompasses the laser tracks. 

\textbf{Pre-processing limitation.} Several data-driven methods use raw infrared images as input without any pre-processing. Raw infrared images are often noisy and high-dimensional, where the informative portion of the data is usually low-dimensional. Examples for the informative portion of the infrared images are the melt pool or the laser tracks, which usually make up only a small region of the infrared images. A lack of pre-processing of the infrared images can cause significant performance degradation of the data-driven methods and also a need for more data~\citep{importance_preprocessing, importance_preprocessing_2}.

On the other hand, the data-driven methods that do use pre-processing often rely on traditional thresholding methods to separate the informative portion of the image from the uninformative portion. Otsu's method is a popular thresholding method, but usually only works well for images with high contrast and low noise~\citep{Otsu, otsu_plus_watershed_monitoring, insitu_monitoring_SLM_zinc_powder, Zinc_successor_paper}. Both of these conditions depend highly on the process parameters and the material used. Furthermore, these methods often cannot distinguish whether the laser is on or off in a given infrared image. Since data-driven methods may only use images that were recorded during scanning, filtering out any other images is crucial. These different limitations can result in restrictions to the end-application and also significantly reduce the efficacy of the data-driven methods.

\textbf{Storage limitation.} In situ measurements produce massive amounts of data and thus, pose a significant challenge for data storage systems~\citep{data_storage_problem, fancy_composite_image, overview_of_off_axis_approaches}. Previous suggestions to overcome this challenge included storing statistical metrics that were calculated from the raw infrared images~\citep{data_storage_problem}. However, reducing high-dimensional data, such as infrared images, to scalar values usually connotes significant information loss and often prevents any meaningful learning for the data-driven methods. Thus, storing data in an effective and efficient manner is desperately needed as the need for data to train the data-driven methods continuously increases.

To address the monitoring, pre-processing, and storage limitations, we study different segmentation algorithms that segment the infrared images into foregrounds and backgrounds. By retaining the informative portion of the infrared images, the foregrounds can be used to \begin{enumerate*}[label=(\alph*)]
\item online monitor the laser tracks,
\item function as pre-processed input for the data-driven methods, and
\item significantly reduce the memory footprint of the off-axis cameras.
\end{enumerate*} We evaluate each algorithm based on segmentation accuracy, computational speed, and spatter detection characteristics. The first two criteria are critical for determining accurate and fast segmentation. The latter is critical to account for ejected spatter that might influence segmentation results as some processes have strong spatter characteristics. By discussing the details and implications of the study results, we provide a guide for selecting suitable segmentation algorithms for LPBF machines to significantly improve process control.

The remainder of the paper is structured as follows: In Section~\ref{section_2}, we define the segmentation problem and provide more detail on the motivation of this paper. We continue by introducing the relevant segmentation algorithms in Section~\ref{section_3} and defining the study methodology in Section~\ref{section_4}. We present our results in Section~\ref{section_5} and discuss how our results can overcome the aforementioned limitations in Section~\ref{section_6}. We conclude by summarizing our findings and addressing future research in Section~\ref{section_7}.
\section{Segmentation definition and motivation} \label{section_2}

In this section, we define the segmentation problem and detail how finding suitable algorithms for this segmentation problem will overcome the aforementioned limitations. The segmentation problem and motivation will guide the selection and study of algorithms in subsequent sections.

\begin{figure}
\centering
\begin{subfigure}[b]{0.65\textwidth}
  \includegraphics[width=\linewidth, trim={0.4cm 5.1cm 4cm 1cm},clip]{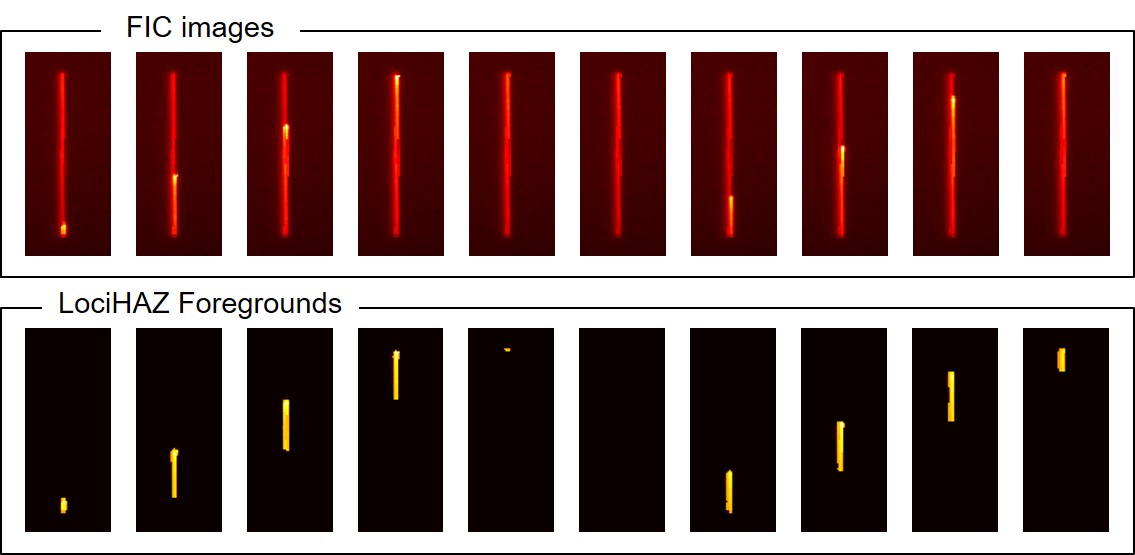}
\caption{Sequence of cropped infrared images}
\label{fig:seq_scanning}
\end{subfigure}
\begin{subfigure}[b]{0.65\textwidth}
\centering
  \includegraphics[width=\linewidth, trim={0.4cm 0.4cm 4cm 5.45cm},clip]{imgs/1_seq_scanning.jpg}
\caption{Manually extracted foregrounds}
\label{fig:seq_scanning_truth}
\end{subfigure}
\caption{Exemplary sequence of cropped infrared images and the corresponding foregrounds. The cropped infrared images are taken from the same SLS build as Figure \ref{fig:full_view_1} but from a different layer. This layer scans the rectangles vertically.}
\end{figure}

Before defining the segmentation problem, we point out several process characteristics relevant to the segmentation problem. The infrared images from off-axis cameras always have a stationary view of the build surface. In other words, the pixel of any image always represents the same location on the build surface. Furthermore, as the laser scans the powder bed (exemplary sequence in Figure~\ref{fig:seq_scanning}), each infrared image contains a snapshot of the laser's effect on the build surface. Some of these scanning effects are new, while others have been captured by previous infrared images. These effects correspond to physical changes on the build surface, such as a different melt pool location and new additions to the laser tracks.

\textbf{Segmentation definition.} Since, in online monitoring and process control, we are often interested in recent changes of a process variable, we define the foreground of each infrared image as follows: the foreground consists of the pixels that correspond to the most recent changes of the laser tracks. Here, most recent refers to the time difference between the 'current' image that was just recorded by the infrared camera and the 'previous' image that was recorded directly before the current image. Conversely, the background consists of all the pixels that do not belong to the foreground. For illustration, Figure~\ref{fig:seq_scanning_truth} shows the foregrounds for the sequence of Figure~\ref{fig:seq_scanning}. We consider the cooldown of previously scanned powder as part of the continuously changing background. 

This definition of the foreground-background segmentation problem offers several advantages that address the monitoring, pre-processing, and storage limitations. 

First, the foregrounds from the segmented output can be used to directly monitor the recently scanned parts of the laser tracks. These parts can be monitored with respect to different properties, such as temperature distribution or geometric dimensions. Furthermore, by creating binary masks from consecutive foregrounds and superimposing the foreground masks, we can extract the pixel values from the current image and monitor all or parts of the previously-scanned laser tracks. This segmentation could provide crucial information about the temperature evolution of the scanned powder. Note that by superimposing foreground masks, we refer to the process of creating one composite mask where a pixel has the value \emph{true} if this pixel was selected as a foreground in at least one of the foregrounds and \emph{false} otherwise.

Second, the foregrounds and the superimposed foreground masks contain most regions of interest, such as the melt pool and the laser tracks. Depending on the process, the foregrounds may also contain the spatter and plume. Since recent data-driven approaches are predominantly focused on monitoring these regions of interest, the data-driven methods can use the foregrounds or modified foregrounds with the superimposed foreground masks as input and eliminate the uninformative portions from the data. This pre-processing can significantly reduce the dimensionality and noise of the data, thus increasing the effectiveness of the data-driven approaches. Furthermore, the segmented outputs also make it easy to eliminate entire images from the input where the laser has not recently scanned the build surface and may not contain any relevant or new information at all. These images will entirely or almost entirely consist of backgrounds.

Finally, based on the flexibility given by the segmented outputs discussed in the previous two paragraphs, the end-user can significantly reduce the memory footprint of the infrared images by storing only the portions of the images that are relevant for the end-application. Since the foregrounds are usually of interest, the LPBF machine can extract the foregrounds from the images and save the raw values for later usage without any information loss. Other portions of the images, such as the previously-scanned powder or possibly the heat-affected zone, can also be stored with the help of the superimposed foreground masks. In the case of the heat-affected zone, the location information of the scanned powder in the infrared images could be used to find the larger heat-affected zone, which directly encompasses the scanned powder. 

\section{Segmentation algorithms} \label{section_3}

In order to address the process limitations mentioned in Section \ref{section_1}, suitable algorithms should meet several criteria. First, the segmentation algorithms should possess high segmentation accuracy while still maintaining low computational complexity. Second, the algorithms should perform well without having to undergo an extensive training phase before deployment and should be applicable to various hardware setups and segmentation purposes. Thus, minimizing human input and avoiding restrictions to the end-applications is crucial. The algorithms may require some calibration of individual parameters, but this calibration should ideally occur only once for similar process setups, such as builds with the same powder material. 

For the scope of this study, we focus on segmentation algorithms provided by the OpenCV and scikit-image libraries~\citep{opencv_library, skimage}. Using OpenCV and scikit-image has several advantages. First, both libraries possess a breadth of segmentation algorithms that use almost no training data and have high potential to provide accurate segmentation with low computational complexity~\citep{opencv_library, openCV_real_time, skimage}. Most of the recently studied methods for the LPBF processes, such as Otsu's method, KNN, and Li's method, are also part of the implemented algorithms ~\citep{otsu_plus_watershed_monitoring, insitu_monitoring_SLM_zinc_powder, Zinc_successor_paper}. Second, most of the OpenCV and scikit-image algorithms are well-tested and optimized for speed, which will allow for comparative analysis of the algorithms' performances~\citep{openCV_real_time, skimage}. Finally, since the algorithms are commonly available and easy to use, suitable algorithms identified in this study can be readily applied to research and commercial machines to address the above process limitations. 

Table~\ref{table:method_overview} lists the algorithms investigated in this study. The methods underlying these algorithms range from thresholding methods, such as basic thresholding or adaptive thresholding, to advanced statistical methods, such as mixture of gaussians~\citep{review_paper_2, comparison_most_cited, opencv_library, skimage}. We provide further detail on these algorithms in the next paragraphs.

\begin{table}[hbt]
    \centering
    \small
    \caption{Overview of algorithms studied in this paper.}
    \captionsetup{justification=centering}
    \begin{tabularx}{0.9\columnwidth} { l X  }
        \toprule
        Name & Description \\
        \midrule
 
        MOG\footnotemark[1]  & Mixture of gaussian algorithm \citep{MOG}\\
        \addlinespace[\var]
        MOG2\footnotemark[1]  & Improved MOG algorithm~\citep{MOG2}\\
        \addlinespace[\var]
        KNN\footnotemark[1]  & K-nearest neighbors algorithm \citep{KNN}\\
        \addlinespace[\var]
        GMG\footnotemark[1]  & {Applies bayesian inference pixel-wise~\citep{GMG}}\\
        \addlinespace[\var]        
        CNT\footnotemark[1]  & {Developed by OpenCV community~\citep{opencv_library}}\\
        \addlinespace[\var]    
        GSOC\footnotemark[1]  & {Developed by Google Summer of Code community~\citep{opencv_library}}\\
        \addlinespace[\var]    
        LSBP\footnotemark[1] & {Uses local SVD patterns~\citep{LSBP}}\\
        \addlinespace[\var]
        Otsu\footnotemark[1]  & {Global, clustering-based method \citep{Otsu, image_binarization_survey}}\\
        \addlinespace[\var]
        Li\footnotemark[2]  & {Global, entropy-based threshold~\citep{Li, image_binarization_survey}}\\
        \addlinespace[\var]
        isodata\footnotemark[2]  & {Global, clustering-based threshold~\citep{isodata, image_binarization_survey}}  \\
        \addlinespace[\var]
        Yen\footnotemark[2]  & {Global, entropy-based threshold ~\citep{Yen, image_binarization_survey}}\\
        \addlinespace[\var]
        Triangle\footnotemark[1]  & {Global threshold using triangle method~\citep{Triangle}}\\
        \addlinespace[\var]
        Sauvola\footnotemark[2]  & {Locally adaptive threshold~\citep{sauvola, image_binarization_survey}}\\
        \addlinespace[\var]
        AdaptMean\footnotemark[1]  & {Local threshold based on adaptive mean~\citep{opencv_library}}\\
        \addlinespace[\var] 
        AdaptGauss\footnotemark[1]  & {Local threshold based on Gaussian weighting~\citep{opencv_library}}\\
        \addlinespace[\var]
        Thresh($\lambda$)  & Global threshold with scalar $\lambda$ \\
        \addlinespace[\var]
        FD  & Frame differencing: Subtracts previous image from current image~\citep{change_detection_review, change_detection_techniques, change_detection_thresholding}\\
        \addlinespace[\var]
        SubMax($\delta$)  & Thresholds image based on $\delta$ and the maximum pixel value of the current image\\
        \addlinespace[\var]   
        FD+'Name'  & FD, then Thresh(1), then method 'Name'  \\
        \addlinespace[\var]
        
        \bottomrule
    \end{tabularx}
    \begin{tablenotes}
        \small
        \item[1] \footnotemark[1] openCV~\citep{opencv_library}
        \item[2] \footnotemark[2] scikit-image~\citep{skimage}
    \end{tablenotes}
    \label{table:method_overview}
\end{table}

The scikit-image algorithms (see footnotes in Table~\ref{table:method_overview}) and several algorithms from OpenCV, specifically the adaptive mean algorithm, the adaptive Gaussian algorithm, and Otsu's method, are thresholding algorithms. These algorithms use either a global or local threshold to determine whether a pixel belongs to the foreground or the background~\citep{image_binarization_reference, image_binarization_survey, image_difference_threshold} Since the laser always increases the powder temperature, and the foreground is directly affected by this increase in temperature, thresholding methods can be an effective way of segmenting an image. Recent work has demonstrated the potential of these segmentation algorithms~\citep{insitu_monitoring_SLM_zinc_powder, Zinc_successor_paper, otsu_plus_watershed_monitoring}.

Most of the OpenCV algorithms belong to the family of background subtraction algorithms that first model the background of an image and then use background subtraction to find the foreground~\citep{bouwman_book, comparison_most_cited, review_paper_2}. OpenCV includes well-established background subtraction algorithms, such as the MOG, MOG2, KNN, and GMG algorithms~\citep{openCV_established_MOG_MOG2}. The MOG and MOG2 algorithms use mixture of gaussians to model the background~\citep{MOG, MOG2}. The KNN algorithm uses a k-nearest neighbors method to update its background model, while the GMG algorithm uses Bayesian inference~\citep{KNN, GMG}. OpenCV also includes recent algorithms, such as the LSBP algorithm~\citep{LSBP}, the GSOC, and the CNT algorithms, where the latter two algorithms do not originate from publications. As demonstrated by previous work, the algorithms strike a balance between computational complexity and segmentation accuracy, and thus, make ideal candidates for solving the segmentation problem~\citep{comparison, review_paper_2, MOG_MOG2_KNN_reference}.

In addition to the algorithms provided by OpenCV and scikit-image, we propose several simple algorithms to either supplement the above algorithms or to establish a baseline for the segmentation study. We detail these algorithms in the next paragraphs.

In order to establish a baseline for the segmentation study, we add two algorithms based on process intuition of the LPBF processes. The first algorithm, called \textit{Thresh($\lambda$)}, uses a constant global threshold $\lambda$  for all images. For each image, the pixels with values above the threshold are selected as foreground, while the remaining pixels are selected as background. The second algorithm, called \textit{SubMax($\delta$)}, selects a pixel as a foreground pixel if its value is close enough to the maximum pixel value of the corresponding image. Here, we define close enough as the pixel values that are only a constant $\delta$ below the maximum pixel value. For instance, if the maximum pixel value for a given image corresponds to $300$~$\degree C$, then only the pixels whose values lay between $(300 - \delta)$~$\degree C$ and $300$~$\degree C$ will be selected as foregrounds. Both of these algorithms exploit the fact that the laser increases the temperature of the build surface by selecting the pixels with the highest intensity values as foregrounds.

Since the foregrounds represent the most recent laser track changes, we use a common change detection technique, called \textit{frame differencing}, to supplement the above algorithms~\citep{change_detection_review, change_detection_techniques, change_detection_thresholding}. Basic frame differencing subtracts the pixel values of the previous image from the pixel values of the current image to detect changes in the current image~\citep{change_detection_techniques}. We denote frame differencing as \textit{FD} and the combination of frame differencing with a subsequent algorithm as \textit{FD+'ID'}, where \textit{'ID'} is the name of the subsequent algorithm (see Table~\ref{table:method_overview}).

For the combination of frame differencing with other algorithms, frame differencing first applies the subtraction to the current image and then supplies the subsequent algorithm with the frame differenced image for further processing. However, before the subsequent algorithm uses the frame differenced image, the image is thresholded with Thresh($1$) to eliminate pixels with near-zero values caused by the frame differencing step. Eliminating these pixels reduces the noise of the images, which can cause significant segmentation problems for the subsequent algorithm. Combining frame differencing with other segmentation algorithms has the potential to improve the segmentation accuracy while avoiding any meaningful increase in computational complexity~\citep{change_detection_thresholding, image_difference_threshold}.

\section{Study methodology} \label{section_4}

In this section, we introduce the study methodology. First, we describe the experimental data and the study procedure used to test the algorithms. Next, we introduce the methods used to evaluate the segmentation accuracy, the computational speed, and the spatter detection. Finally, we provide detail on the computational hardware and describe how each algorithm's parameters are tuned to ensure an adequate comparison between the algorithms. 

\subsection{Study setup}

Two different research machines were used to collect data for this study; one SLS machine and one SLM machine. The SLS machine is located at the University of Texas at Austin, while the SLM machine is located at the Rensselaer Polytechnic Institute. Both machines are equipped with an off-axis infrared camera; the SLS machine uses a FLIR 6701sc camera with a resolution of 640 by 512 at a frame rate of 60 \si{Hz}. The SLM machine uses a FLIR A320 camera with an effective resolution of 50 by 50 at a frame rate of 60 \si{Hz}. 

\begin{figure}
\centering
\begin{subfigure}{.35\textwidth}
  \centering
  \includegraphics[height=0.85\linewidth, trim={2.1cm 2cm 4.3cm 1.5cm},clip]{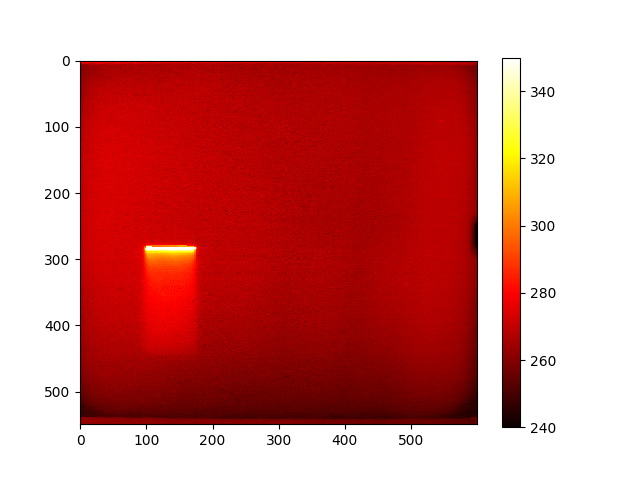}
  \caption{SLS off-axis infrared image.}
  \label{fig:first_sub1}
\end{subfigure}
\begin{subfigure}{.35\textwidth}
  \centering
  \includegraphics[height=0.85\linewidth]{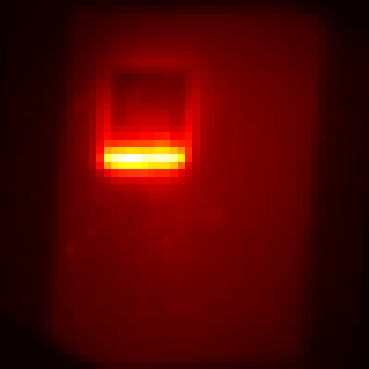}
  \caption{SLM off-axis infrared image.}
  \label{fig:first_sub2}
\end{subfigure}
\caption{Exemplary infrared images recorded during the processing of the first rectangular cross-section for the SLS and SLM builds. The images are from the same layer as Figure~\ref{fig:full_view_off_axis}.}
\label{fig:first_rectangle_finished_scanning}
\end{figure}

Both machines produced four rectangular cuboids. The SLS machine used carbon fiber reinforced polyether ether ketone (PEEK), while the SLM machine used cobalt-chrome. We show exemplary infrared images from both builds in Figure~\ref{fig:first_rectangle_finished_scanning}. These images stem from the same layer as Figure~\ref{fig:full_view_off_axis}. The SLS images have been transformed to account for the non-perpendicular angle of view to the build surface. The SLM images have not been modified.  

For this study, we select one batch of images from the SLM dataset and two batches of images from the SLS dataset. Each batch was recorded during the scanning of the first rectangular cross-section. Since some algorithms need a few images to initialize, each batch starts with roughly 50 images before the laser starts scanning. To imitate conditions in online monitoring and process control, we sequentially apply the infrared images from each batch to the algorithms. After the algorithms segment the images, the images are evaluated without any post-processing.

\subsection{Segmentation accuracy}

In order to evaluate each algorithm's segmentation accuracy, we test the algorithms on the SLS dataset and compare the segmented outputs to the segmentation ground truth. To determine the segmentation ground truth, we can use insights from the process setup of the SLS build. We know that the laser processed the rectangular cross-section with 215 parallel and overlapping scan lines, where each scan line was processed with the same process parameters. The processed scan lines show up as horizontal laser tracks on the infrared images. Based on this information, we can use the average width of a laser track and the approximate location of the laser to determine the segmentation ground truth for each image.

Finding the laser's approximate location in each image is straightforward since the highest temperatures on the build surface occur close to where the laser spot is. Thus, the location can be determined by the pixels with the highest intensity values in each image. Finding the average width of the laser tracks is more challenging. First, we need to quantify what belongs to a laser track. Second, we need to determine the average width from multiple single laser tracks.

Since the intensity distribution of the laser spot resembles a Gaussian profile, the temperature transition from a single laser track to the surrounding heat-affected zone and unprocessed powder is mostly smooth on the infrared images. To determine which pixels belong to a single laser track, we define a cutoff temperature. Any pixels with intensity values above the cutoff temperature belong to this laser track. This cutoff temperature needs to be sufficiently above the build surface temperature to ensure that the elevated intensity values are only caused by the direct scanning of the laser.  

For this study, we define the cutoff temperature as $T_{c} = 3 \cdot \sigma_{bs} + T_{max},$ where $T_{max}$ is the maximum temperature of the build surface and $\sigma_{bs}$ is the standard deviation for the temperature distribution of the build surface. $\sigma_{bs}$ and $T_{max}$ are calculated from images that show the build surface before the laser starts scanning. In the case of the selected batches for the SLS build, the average $\sigma_{bs}$ is around $6~\degree C$, while the average $T_{max}$ is around $277~\degree C$. We, thus, set $T_{c}$ to $295~\degree C$. Based on this cutoff temperature, we can determine the average width of a single laser track by counting the average amount of pixels stacked on top of each other along the length of the laser track. Figure~\ref{fig:first_scan_line_ROI} shows an exemplary laser track that is used for estimating the width, where the width is in the vertical direction of the image. We use multiple single laser tracks recorded during the beginning of different cross-sections and average over the measured widths.

Given the average laser track width and the laser's approximate location for each image, we can find the most recent laser track changes as follows: We know that each laser track is scanned from left to right along the horizontal axis. Thus, we approximate the most recent laser track changes with a rectangular box along the horizontal axis. This box has the same width in the vertical direction as the average laser track width. Furthermore, the box stretches from the laser location of the current image to the laser location of the previous image. If the current image does not have a laser location, e.g., because the laser is off, the box sits between the previous laser location and the right border of the rectangular cross-section. If the previous image does not have a laser location, then the box sits between the current laser location and the left border of the rectangular cross-section. If neither the current nor the previous image have laser locations, then no box is drawn as no foreground is present in the current image.

\begin{figure}
\centering
  \includegraphics[width=0.7\columnwidth, trim={2cm 4cm 2cm 4cm},clip]{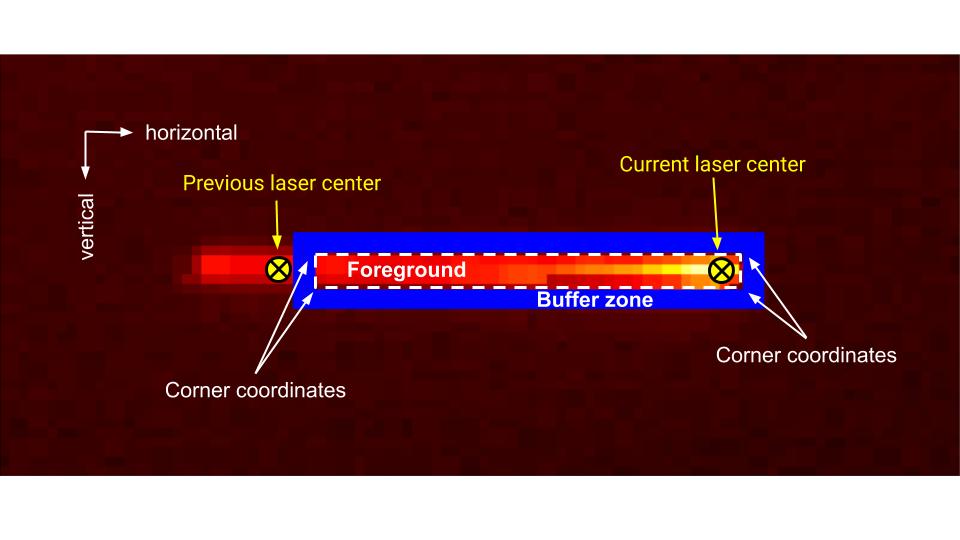}
\caption{Single laser track annotated with the foreground ground truth and the corresponding buffer zone.}
\label{fig:first_scan_line_annotated}
\end{figure}

To demonstrate how the exact box location for each image is determined, we use the same laser track image from Figure~\ref{fig:first_scan_line_ROI} and annotate the image in Figure~\ref{fig:first_scan_line_annotated}. For the vertical direction, the box coordinates are exactly half the average laser track width above and below the vertical coordinate of the current laser center. Since the laser has a circular beam spot on the build surface, the horizontal coordinates of the right side corners are half the average laser track width to the right of the current laser center. For the left side corners, the horizontal coordinates are to the right of the previous laser center location. However, the horizontal distance between the left side corner coordinates and the previous laser center location is half the average laser track width plus an extra distance due to a buffer zone (blue box in Figure~\ref{fig:first_scan_line_annotated}).

\begin{figure}
\centering
\begin{subfigure}{.35\textwidth}
  \centering
  \includegraphics[height=0.85\linewidth, trim={2.1cm 2cm 4.3cm 1.5cm},clip]{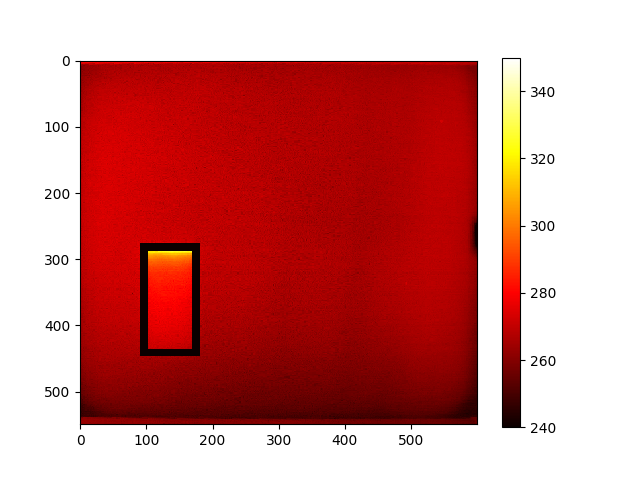}
  \caption{SLS off-axis infrared image.}
  \label{fig:mask_sub1}
\end{subfigure} 
\begin{subfigure}{.35\textwidth}
  \centering
  \includegraphics[height=0.85\linewidth]{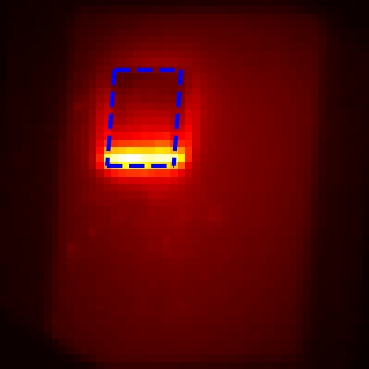}
  \caption{SLM off-axis infrared image.}
  \label{fig:mask_sub2}
\end{subfigure}
\caption{Annotated infrared images from Figure~\ref{fig:first_rectangle_finished_scanning}. The buffer zone for the SLS cross-section is marked in black. The contour of the SLM cross-section is marked with dashed lines in blue.}
\label{fig:exemple_masks}
\end{figure}

To eliminate any dependency of the study results on the cutoff temperature, we use a small buffer zone around each laser track that will not be part of the segmentation evaluation. This buffer zone eliminates the pixels representing the transition zone between the laser track and the surrounding areas, such as the heat-affected zone and the unprocessed powder. Using a buffer zone allows the end-user to use lower cutoff temperatures if desired. Furthermore, we use an outer buffer zone around the entire rectangular cross-section to eliminate any segmentation uncertainties at the transition of the processed powder to the unprocessed powder (see Figure~\ref{fig:mask_sub1}).

Given the box coordinates, we can determine the segmentation ground truth for each infrared image. The pixels within each box represent the current foreground, while the pixels outside of the box represent the background. Before the segmented outputs are compared to the ground truth, the inner and outer buffer zones eliminate the pixels representing the transition zones from the evaluation. For this study, the border thickness of the inner buffer zone is set to half the average laser track width, while the thickness for the outer buffer zone is set to five times the average laser track width.

We can evaluate the segmentation accuracy with the commonly-applied F\textsubscript{1}-score. By comparing the binary mask of each segmentation output pixel-wise to the corresponding ground truth image, we can calculate the true positive rate $TP$, false positive rate $FP$, true negative rate $TN$ and false negative rate $FN$. From these rates, we can calculate the resulting F\textsubscript{1}-score by \[ F_{1} = 2 \cdot \frac{Precision \cdot Recall}{Precision + Recall}\] where $Precision$ and $Recall$ are defined as \[Precision = \frac{TP}{TP+FP}, \quad Recall = \frac{TP}{TP+FN}.\]

\subsection{Parameter tuning}

Several of the introduced algorithms have adjustable parameters that can significantly differ the segmentation accuracy of each algorithm. To ensure adequate comparison of the different algorithms, we test each algorithm twice; once with default parameters and once with calibrated parameters. We use a small number of images from the other SLS batch that were recorded during the scanning of the first 20 consecutive laser tracks to tune each algorithm's parameters. By applying a random grid search to test hundreds of different sets of parameters on these laser tracks and scoring each set with the F\textsubscript{1}-score, we calibrate each algorithm's parameters.

Table~\ref{table:parameters} in~\ref{appendix_tables} shows the default and calibrated parameters for each algorithm. The default for the Thresh($\lambda$) parameter was set to the calculated $T_c$ of $295$, while the default for the FD+Thresh($\lambda$) was set to 3. The default for the SubMax($\delta$) parameter was set to 20.

\subsection{Computational speed}

We test the computational speed of each algorithm by timing the sequential processing on a batch of 2000 images from the SLS dataset and calculating the average computation speed for each image. The algorithms are implemented in Python but are often speed-optimized with underlying C or C++ implementations by OpenCV and scikit-image~\citep{opencv_library, openCV_real_time, skimage}. For the algorithms we implemented, e.g., the frame differencing method, we rely on numpy~\citep{numpy}. To ensure a relative comparison between the different algorithms, we use the CPU implementation and the algorithm's calibrated parameters. We perform our computation on a laptop with an Intel i9-9750H six core (2.6 up to 4.5 GHz, 12 MB Cache) and 32 GB DDR4 RAM.

\subsection{Spatter detection}

Spatter is often present during the processing of metal powder. Spatter can be part of the monitoring variables, as demonstrated by previous work, or can be seen as a disturbance~\citep{segmentation_2, segmentation_4}. In the case of this segmentation study, the spatter can cause false positives and create foregrounds that would suggest changes to nonexistent laser tracks. The testing of the segmentation accuracy with the SLS dataset circumvented this problem since PEEK does not create noticeable spatters during the processing of the build surface. However, the segmentation accuracy may change drastically if the algorithms are applied to processes with high amounts of spatter ejection. Thus, we separately investigate the spatter detection for the segmentation algorithms on the SLM dataset. We use the SLM dataset due to its strong spatter ejection characteristics as the spatter was repeatedly ejected across the entire build surface and significantly obscured the laser tracks.

The upper-left rectangular cross-section is the first cross-section scanned for each layer of the SLM build. We use the first 400 images of a sample layer that show the processing of the upper-left cross-section and apply the algorithms to the images. We manually analyze the sequence of images for each algorithm by qualitatively looking for obvious artifacts in the foregrounds. If a given algorithm does not detect spatter in its foregrounds, the foregrounds would only exhibit a single coherent laser track, such as Figure \ref{fig:seq_scanning_truth}. Otherwise, the foreground will include additional nonzero pixels across the entire image.

To support and illustrate our manual analysis, we use a binary composite image for each algorithm that shows which pixels have been part of a foreground at least once and which pixels were never selected as part of a foreground. The binary composite image is created by first turning the foregrounds of a given algorithm into binary masks, where the foreground pixels have the value \textit{one} and the background pixels the value \textit{zero}. Following, the masks are summed up pixel-wise. The resulting binary composite image has nonzero values for pixels that have been part of a foreground at least once and pixels with zero values that were never selected as part of a foreground. If a given algorithm includes spatter in its foregrounds, the binary composite image would show numerous pixels selected as foreground that lay outside of the upper-left rectangular cross-section. 

For visualization purposes, we manually extract the approximate border contour of the upper-left cross-section (Figure~\ref{fig:mask_sub2}) and project the contour onto each binary composite image. The area of the nonzero pixels in the binary composite image should approximately match the area represented by the projected contour. Some nonzero pixels fall slightly outside of the contour since this 'overflow' of nonzero pixels around the contour is caused by the difference in segmentation of the transition zone between the laser tracks and the unprocessed powder. 

For the segmentation of the SLM batch, we use the default parameters for the segmentation algorithms. The calibrated parameters from the SLS dataset may not translate to the SLM dataset since the pixel intensity values from the SLS dataset are different from the SLM dataset.
\section{Results} \label{section_5}

\subsection{Segmentation accuracy} \label{Seg_acc}

\begin{table}[phtb]
    \centering
    \small
    \caption{Segmentation accuracy of tested algorithms on the SLS dataset. Algorithms are ordered by calibrated F\textsubscript{1}-score. The bold algorithm is one of the baseline algorithms.}
    \captionsetup{justification=centering}
    \begin{tabular}{llr}
        \toprule
        Algorithms & Parameters & F\textsubscript{1}-score \\
        \midrule
        \multirow{2}{*}{FD+Thresh($\lambda$)}        & calibrated            & 0.93 \\
        {}                                  & default               & 0.88 \\
        \addlinespace[5pt]
        
        \multirow{2}{*}{FD+GSOC}               & calibrated            & 0.84\\
        {}                                  & default               & 0.41 \\
        \addlinespace[5pt]   
        
        FD+Yen                                 & \multicolumn{1}{c}{--}            & 0.78\\
        \addlinespace[5pt]
        
        \multirow{2}{*}{FD+CNT}                & calibrated            & 0.70 \\
        {}                                  & default               & 0.03 \\
        \addlinespace[5pt]         
        
        \multirow{2}{*}{FD+MOG}                & calibrated            & 0.67 \\
        {}                                  & default               & 0.60 \\
        \addlinespace[5pt]                
        
        \multirow{2}{*}{\textbf{Thresh}}               & calibrated            & 0.62\\
        {}                                  & default               & 0.17 \\
        \addlinespace[5pt]   
        
        FD+isodata                             & \multicolumn{1}{c}{--}            & 0.61\\
        \addlinespace[5pt]
        
        FD+Otsu                           & \multicolumn{1}{c}{--}            & 0.60 \\
        \addlinespace[5pt]
                
        FD+Triangle                                & \multicolumn{1}{c}{--}            & 0.60 \\
        \addlinespace[5pt]                
        
        \multirow{2}{*}{GSOC}               & calibrated            & 0.60\\
        {}                                  & default               & 0.13 \\
        \addlinespace[5pt]        
        
        \multirow{2}{*}{MOG}                & calibrated            & 0.58 \\
        {}                                  & default               & 0.35 \\
        \addlinespace[5pt]
        
        \multirow{2}{*}{CNT}                & calibrated            & 0.56\\
        {}                                  & default               & 0.04 \\
        \addlinespace[5pt]     
        
        \multirow{2}{*}{FD+Sauvola}                & calibrated            & 0.54\\
        {}                                  & default               & 0.48 \\
        \addlinespace[5pt]  
        
        FD+Li                                  & \multicolumn{1}{c}{--}            &   0.52\\
        \addlinespace[5pt]
        
        \multirow{2}{*}{FD+GMG}                & calibrated            & 0.38 \\
        {}                                  & default               & 0.36 \\
        \addlinespace[5pt]        
        
        \multirow{2}{*}{FD+MOG2}                & calibrated            & 0.35 \\
        {}                                  & default               & 0.33 \\
        \addlinespace[5pt]

        \multirow{2}{*}{GMG}                & calibrated            & 0.29\\
        {}                                  & default               & 0.29 \\
        \addlinespace[5pt]        
        
        \multirow{2}{*}{LSBP}                 & calibrated            & 0.29 \\
        {}                                  & default               & 0.26 \\
        \addlinespace[5pt]
        
        \multirow{2}{*}{MOG2}               & calibrated            & 0.29\\
        {}                                  & default               & 0.17 \\
        \addlinespace[5pt]        
        
        \multirow{2}{*}{FD+KNN}                & calibrated            & 0.22 \\
        {}                                  & default               & 0.21 \\
        \addlinespace[5pt]           

        \multirow{2}{*}{KNN}                & calibrated            & 0.14 \\
        {}                                  & default               & 0.10 \\
        \addlinespace[5pt]
        
        Triangle                            & \multicolumn{1}{c}{--}            & 0.02 \\
        \addlinespace[5pt]
        
        Yen                            & \multicolumn{1}{c}{--}            & 0.02 \\
        \addlinespace[5pt]
        
        \multirow{2}{*}{FD+LSBP}                 & calibrated            & 0.01 \\
        {}                                  & default               & 0.01 \\
        \addlinespace[5pt]        
        
        Otsu                            & \multicolumn{1}{c}{--}            & 0.01 \\
        \addlinespace[5pt]        
        
        Others                  & \multicolumn{1}{c}{--}            & 0.00 \\
        \addlinespace[5pt]

        \bottomrule
    \end{tabular}
    \label{table:F1_score}
\end{table}
\begin{table}[phtb]
    \centering
    \small
    \caption{Average computation times of tested algorithms. Algorithms are ranked by time in increasing order.}
    \captionsetup{justification=centering}
    \begin{tabular}{lc}
        \toprule
        ID & Time per image [ms] \\
        \midrule
        Thresh($\lambda$)  & 0.21 \\
        \addlinespace[\var]
        FD & 0.29 \\
        \addlinespace[\var]
        FD+Thresh($\lambda$) & 0.47 \\
        \addlinespace[\var]
        Triangle   & 0.69  \\
        \addlinespace[\var]
        Otsu   & 0.72\\
        \addlinespace[\var]
        CNT   & 0.74\\
        \addlinespace[\var]    
        SubMax($\delta$)  & 0.96 \\
        \addlinespace[\var]
        FD+Triangle   & 1.09 \\
        \addlinespace[\var]
        Yen   & 1.10\\
        \addlinespace[\var]
        isodata   & 1.12  \\
        \addlinespace[\var]
        FD+Otsu   & 1.15 \\
        \addlinespace[\var]
        AdaptMean   & 1.21\\
        \addlinespace[\var]
        FD+AdaptMean   & 1.40\\
        \addlinespace[\var]
        FD+isodata   & 1.50 \\
        \addlinespace[\var]
        FD+Yen   & 1.51 \\
        \addlinespace[\var]
        FD+AdaptGauss   & 1.66\\
        \addlinespace[\var]
        FD+CNT   & 1.67\\
        \addlinespace[\var]
        MOG2   & 2.31\\
        \addlinespace[\var]
        KNN   & 2.59\\
        \addlinespace[\var]
        FD+MOG2   & 3.32\\
        \addlinespace[\var]
        MOG   & 5.17\\
        \addlinespace[\var]
        FD+KNN   & 5.19\\
        \addlinespace[\var]
        FD+MOG   & 5.33\\
        \addlinespace[\var]
        FD+Li   & 7.02 \\
        \addlinespace[\var]
        Sauvola   & 8.36\\
        \addlinespace[\var]
        GMG   & 8.44\\
        \addlinespace[\var]
        Li   & 9.21\\
        \addlinespace[\var]
        FD+GMG   & 9.39\\
        \addlinespace[\var]
        AdaptGauss   & 15.77\\
        \addlinespace[\var]
        FD+LSBP   & 16.00\\
        \addlinespace[\var]
        LSBP   & {16.83}\\
        \addlinespace[\var]    
        FD+Sauvola   & 23.74\\
        \addlinespace[\var]
        GSOC   & 267.37\\
        \addlinespace[\var] 
        FD+GSOC   & 296.03\\
        \addlinespace[\var]

        \bottomrule
    \end{tabular}
    \label{table:speed}
\end{table}

The simple combination algorithm FD+Thresh($\lambda$) performed best on the segmentation accuracy test (Table~\ref{table:F1_score}). Even with default parameters, FD+Thresh($\lambda$) had a better F\textsubscript{1}-score than all of the other algorithms. Several algorithms, such as FD+GSOC and FD+Yen, trailed FD+Thresh($\lambda$) but still performed better than the baseline algorithm Thresh($\lambda$). All algorithms that performed better than Thresh($\lambda$) were combination algorithms that used the frame differencing method as the first processing step. 

Most of the OpenCV algorithms that were not combination algorithms scored in the midfield of the F\textsubscript{1}-score. Most of the thresholding algorithms, on the other hand, scored at the bottom of the F\textsubscript{1}-score. Some of these thresholding algorithms, such as the Yen, isodata, and Sauvola algorithms, did not score higher than $0.00$ and are grouped as \textit{Others} in Table~\ref{table:F1_score}. This group also includes the other baseline algorithm SubMax($\delta$). For a detailed breakdown of each algorithm's performance that includes the precision and recall, we refer the reader to Table~\ref{table:all_results1} and Table~\ref{table:all_results2} in \ref{appendix_tables}.

\begin{figure*}[p]
    \centering 
    
\begin{subfigure}{\textwidth}
  \includegraphics[width=0.115\linewidth]{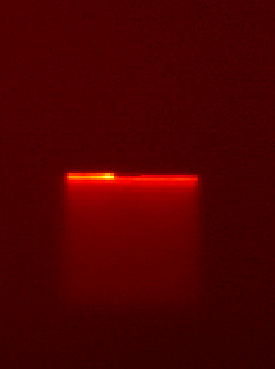}
    \hfil 
  \includegraphics[width=0.115\linewidth]{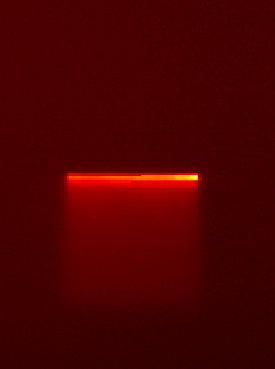}
    \hfil 
  \includegraphics[width=0.115\linewidth]{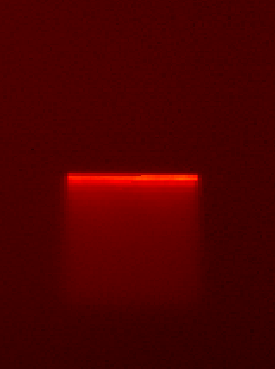}
      \hfil 
  \includegraphics[width=0.115\linewidth]{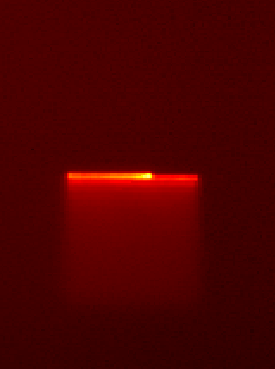}
    \hfil 
  \includegraphics[width=0.115\linewidth]{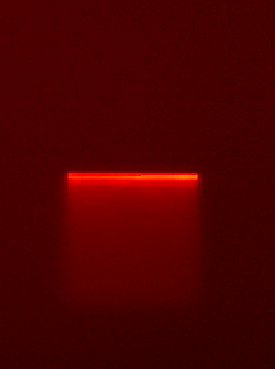}
      \hfil 
  \includegraphics[width=0.115\linewidth]{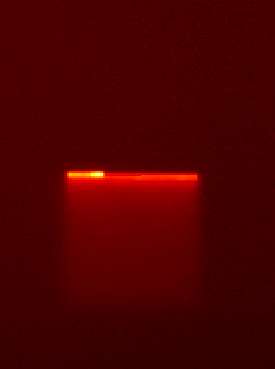}
    \hfil 
  \includegraphics[width=0.115\linewidth]{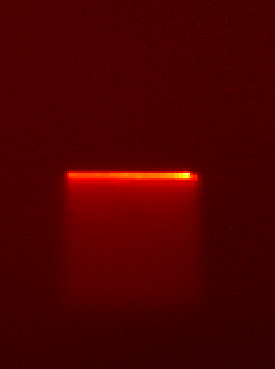}
\subcaption[]{Cropped raw infrared images from the SLS build (images from the same layer as Figure \ref{fig:mask_sub1})}
\end{subfigure}
\medskip

\begin{subfigure}{\textwidth}
  \includegraphics[width=0.115\linewidth]{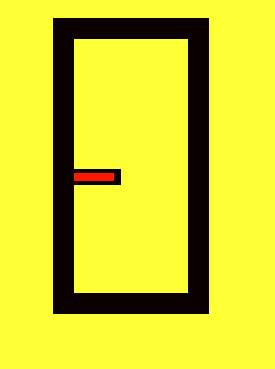}
    \hfil 
  \includegraphics[width=0.115\linewidth]{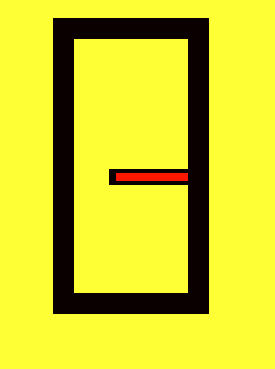}
    \hfil 
  \includegraphics[width=0.115\linewidth]{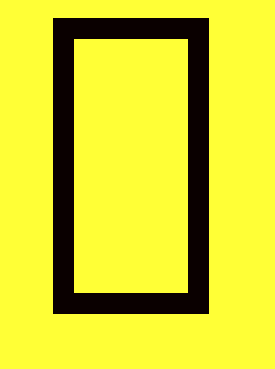}
      \hfil 
  \includegraphics[width=0.115\linewidth]{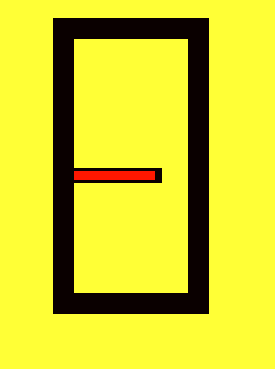}
    \hfil 
  \includegraphics[width=0.115\linewidth]{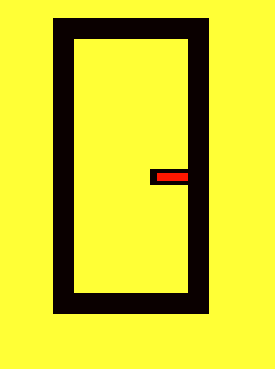}
      \hfil 
  \includegraphics[width=0.115\linewidth]{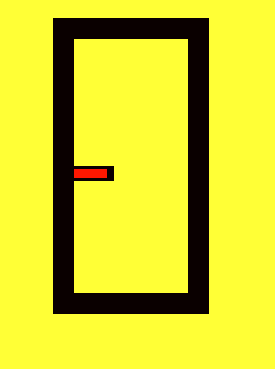}
    \hfil 
  \includegraphics[width=0.115\linewidth]{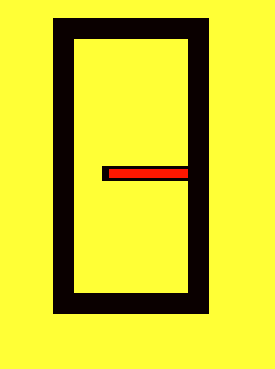}
\subcaption[]{Segmentation ground truth. Black marks the buffer zone, red the foreground, and yellow the background.}   
\end{subfigure}
\medskip

\begin{subfigure}{\textwidth}
  \includegraphics[width=0.115\linewidth]{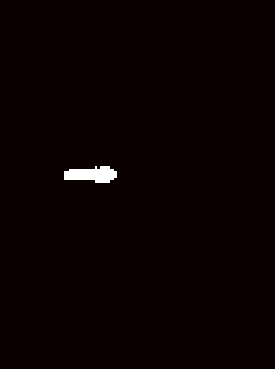}
    \hfil 
  \includegraphics[width=0.115\linewidth]{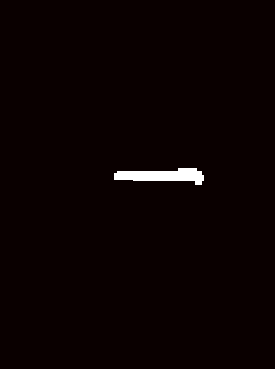}
    \hfil 
  \includegraphics[width=0.115\linewidth]{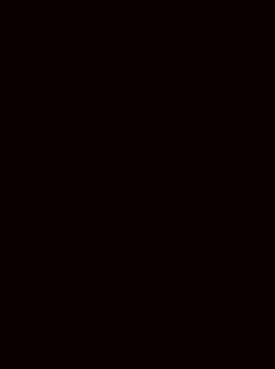}
      \hfil 
  \includegraphics[width=0.115\linewidth]{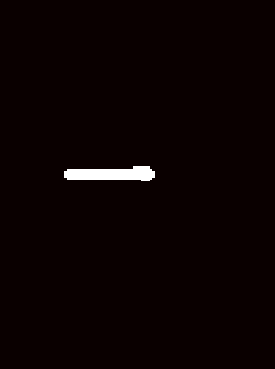}
    \hfil 
  \includegraphics[width=0.115\linewidth]{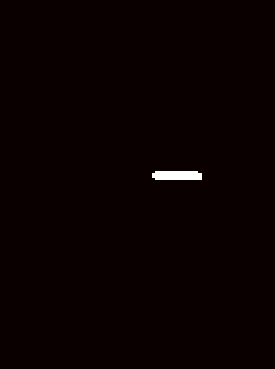}
      \hfil 
  \includegraphics[width=0.115\linewidth]{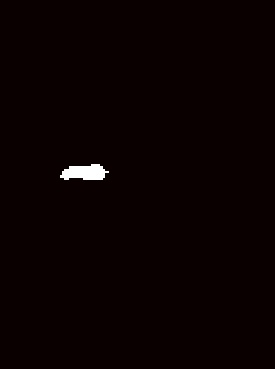}
    \hfil 
  \includegraphics[width=0.115\linewidth]{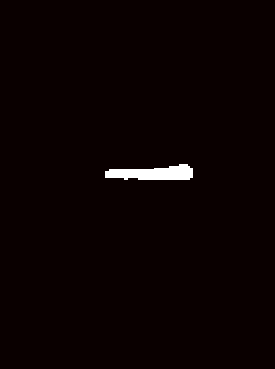}
\subcaption[]{FD + Thresh($\lambda$)}   
\end{subfigure}

\medskip

\begin{subfigure}{\textwidth}
  \includegraphics[width=0.115\linewidth]{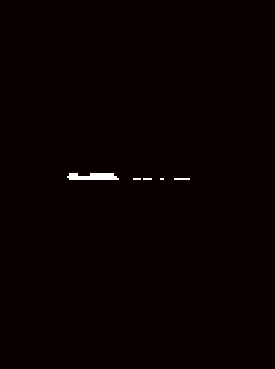}
    \hfil 
  \includegraphics[width=0.115\linewidth]{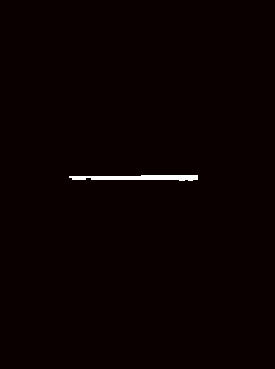}
    \hfil 
  \includegraphics[width=0.115\linewidth]{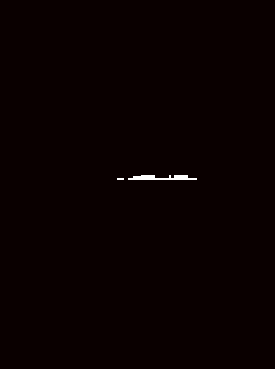}
      \hfil 
  \includegraphics[width=0.115\linewidth]{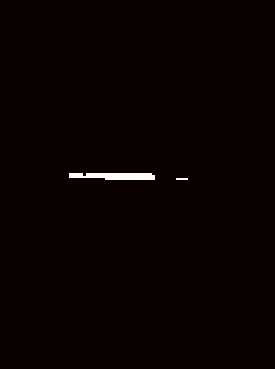}
    \hfil 
  \includegraphics[width=0.115\linewidth]{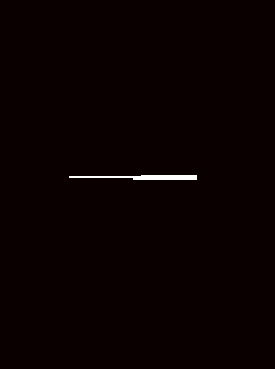}
      \hfil 
  \includegraphics[width=0.115\linewidth]{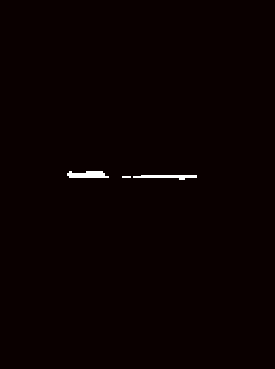}
    \hfil 
  \includegraphics[width=0.115\linewidth]{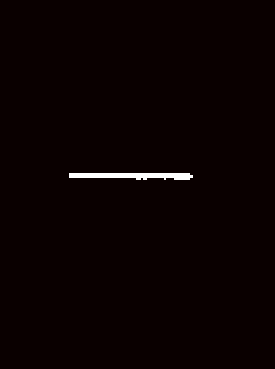}
\subcaption[]{Thresh($\lambda$)}
\end{subfigure}

\medskip

\begin{subfigure}{\textwidth}
  \includegraphics[width=0.115\linewidth, frame]{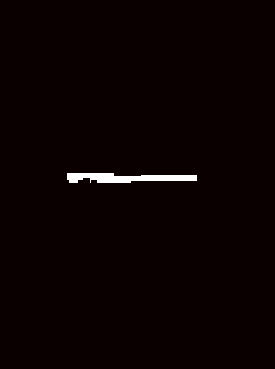}
    \hfil 
  \includegraphics[width=0.115\linewidth, frame]{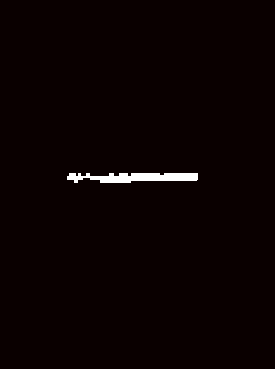}
    \hfil 
  \includegraphics[width=0.115\linewidth, frame]{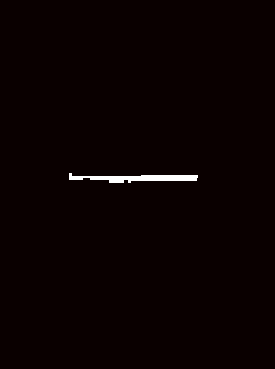}
      \hfil 
  \includegraphics[width=0.115\linewidth, frame]{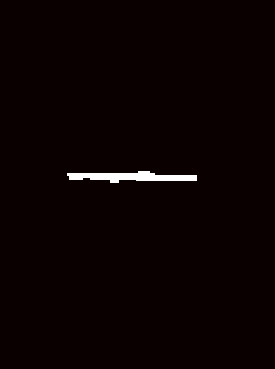}
    \hfil 
  \includegraphics[width=0.115\linewidth, frame]{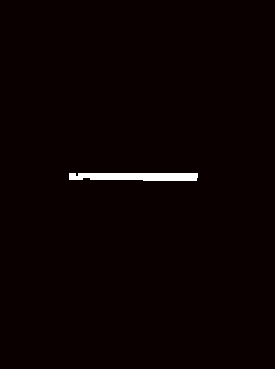}
      \hfil 
  \includegraphics[width=0.115\linewidth, frame]{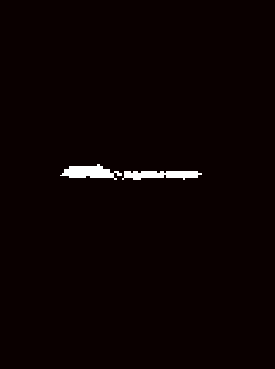}
    \hfil 
  \includegraphics[width=0.115\linewidth, frame]{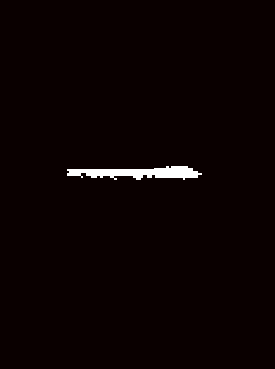}
\subcaption[]{MOG}
\end{subfigure}

\medskip
\begin{subfigure}{\textwidth}
  \includegraphics[width=0.115\linewidth, frame]{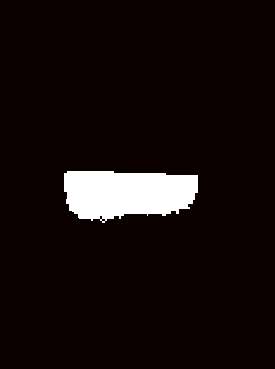}
    \hfil 
  \includegraphics[width=0.115\linewidth, frame]{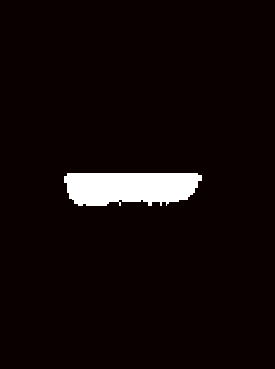}
    \hfil 
  \includegraphics[width=0.115\linewidth, frame]{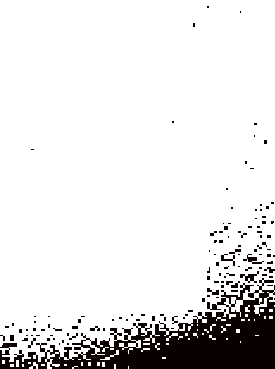}
      \hfil 
  \includegraphics[width=0.115\linewidth, frame]{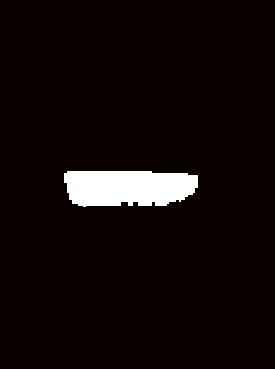}
    \hfil 
  \includegraphics[width=0.115\linewidth, frame]{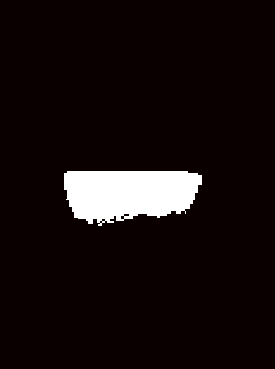}
      \hfil 
  \includegraphics[width=0.115\linewidth, frame]{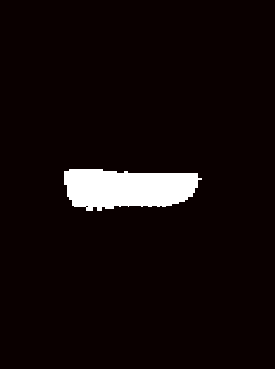}
    \hfil 
  \includegraphics[width=0.115\linewidth, frame]{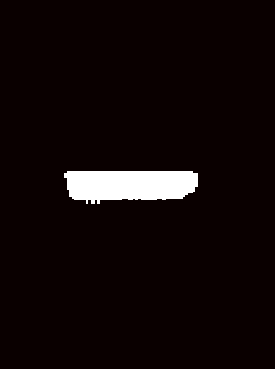}
\subcaption[]{Otsu's method}
\end{subfigure}

\caption{Segmentation results of different algorithms for an exemplary sequence of infrared images. The images are cropped to the area of the first cross-section.}
\label{fig:segment_results}

\end{figure*}

To demonstrate the segmentation results, we present the results from different algorithms for an exemplary sequence of the SLS batch in Figure~\ref{fig:segment_results}. The first two rows show the raw infrared images and the corresponding ground truth images. The ground truth images mark the current foreground as red, the current background as yellow, and the current buffer zones as black. The subsequent rows show the binary masks of the segmentation results. The white pixels represent the foreground, while the black pixels represent the background. All of the images are cropped to the area of the first cross-section (see Figure ~\ref{fig:first_sub1} and ~\ref{fig:mask_sub1}).

As indicated by the third row of Figure~\ref{fig:segment_results}, FD+Thresh($\lambda$) successfully detected the desired foreground pixels. FD+Thresh also recognized when there are no foreground pixels present due to the laser not scanning the build surface (see third column). In contrast, Thresh, MOG, and Otsu's method, did not successfully segment the images into foregrounds and backgrounds. All three algorithms exhibit foreground pixels that span the entire horizontal length of the cross-section and include false positives during the time the laser is off.

\subsection{Computational speed}

As indicated by Table~\ref{table:speed}, the Thresh($\lambda$), FD, and FD+Thresh($\lambda$) algorithms achieved the fastest average computation time per image due to their low computational complexity. The thresholding algorithms Triangle and Otsu as well as the CNT and SubMax($\delta$) algorithms also demonstrated fast computation with sub-millisecond times. The remaining algorithms took anywhere between 1 to 25 milliseconds to process each image. The only algorithms that required considerably more time were the GSOC algorithm and its combination algorithm FD+GSOC. The computational speed of these two algorithms was an order of magnitude higher than any other algorithm.

For some of the thresholding algorithms, the speed of the individual thresholding algorithm was higher than the speed of the corresponding combination algorithm. Since the thresholding step in the combination algorithm eliminates some of the nonzero pixels, the decrease in speed is likely due to a smaller number of nonzero pixels present in each image.

\subsection{Spatter detection}

\begin{figure*}[htb!]
    \centering 
\begin{subfigure}{0.17\textwidth}
  \includegraphics[width=0.95\linewidth, frame]{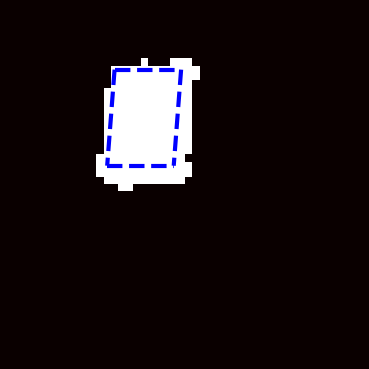}
  \caption{MOG\textsuperscript{1}}
  \label{fig:1}
\end{subfigure}\hfil 
\begin{subfigure}{0.17\textwidth}
  \includegraphics[width=0.95\linewidth, frame]{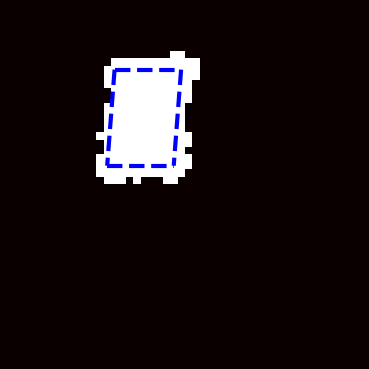}
  \caption{MOG2\textsuperscript{1}}
  \label{fig:2}
\end{subfigure}\hfil 
\begin{subfigure}{0.17\textwidth}
  \includegraphics[width=0.95\linewidth, frame]{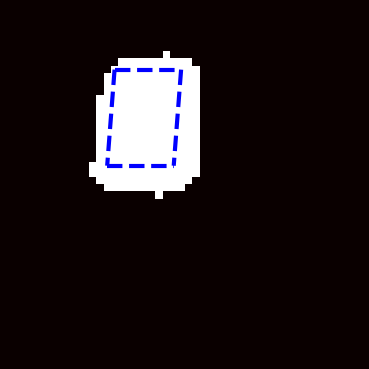}
  \caption{CNT\textsuperscript{1}}
  \label{fig:3}
\end{subfigure}\hfil 
\begin{subfigure}{0.17\textwidth}
  \includegraphics[width=0.95\linewidth, frame]{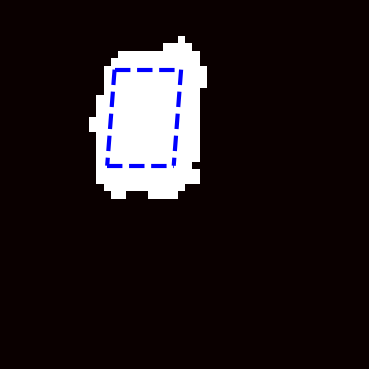}
  \caption{GSOC\textsuperscript{1}}
  \label{fig:4}
\end{subfigure}\hfil 
\begin{subfigure}{0.17\textwidth}
  \includegraphics[width=0.95\linewidth, frame]{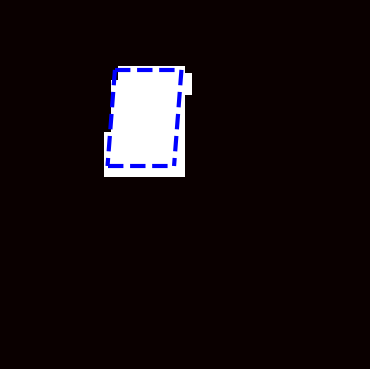}
  \caption{Thresh($\lambda$)\textsuperscript{1}}
  \label{fig:10}
\end{subfigure}\hfil 
\medskip
\begin{subfigure}{0.17\textwidth}
  \includegraphics[width=0.95\linewidth, frame]{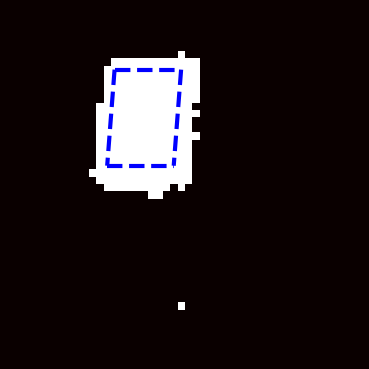}
  \caption{KNN\textsuperscript{1}}
  \label{fig:5}
\end{subfigure}\hfil 
\begin{subfigure}{0.17\textwidth}
  \includegraphics[width=0.95\linewidth, frame]{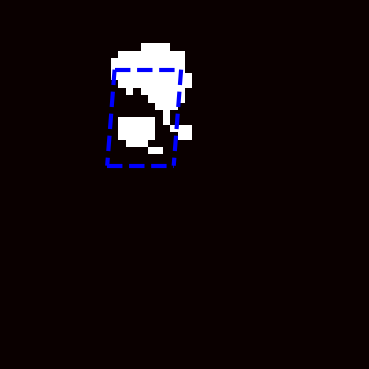}
  \caption{GMG\textsuperscript{2}}
  \label{fig:8}
\end{subfigure}\hfil
\begin{subfigure}{0.17\textwidth}
  \includegraphics[width=0.95\linewidth, frame]{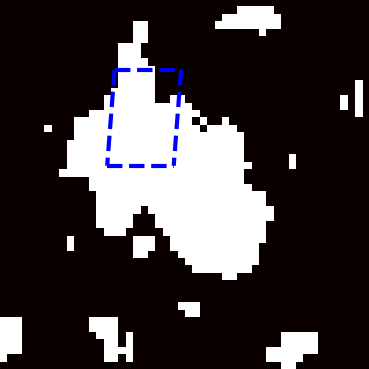}
  \caption{FD + GMG\textsuperscript{2}}
  \label{fig:9}
\end{subfigure}\hfil 
\begin{subfigure}{0.17\textwidth}
  \includegraphics[width=0.95\linewidth, frame]{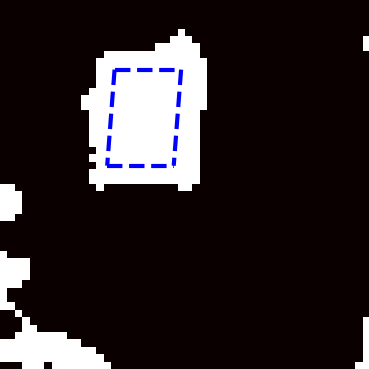}
  \caption{LSBP\textsuperscript{2}}
  \label{fig:7}
\end{subfigure}\hfil 
\begin{subfigure}{0.17\textwidth}
  \includegraphics[width=0.95\linewidth, frame]{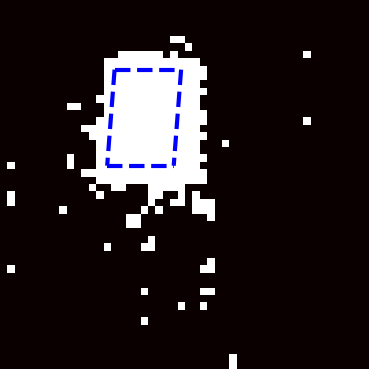}
  \caption{FD + Thresh($\lambda$)\textsuperscript{2}}
  \label{fig:6}
\end{subfigure}\hfil 

\medskip

\begin{subfigure}{0.17\textwidth}
  \includegraphics[width=0.95\linewidth, frame]{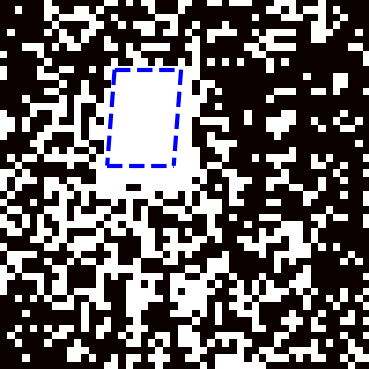}
  \caption{FD + MOG2\textsuperscript{2}}
  \label{fig:11}
\end{subfigure}\hfil 
\begin{subfigure}{0.17\textwidth}
  \includegraphics[width=0.95\linewidth, frame]{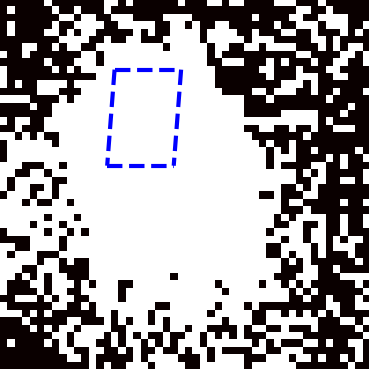}
  \caption{FD + Yen\textsuperscript{2}}
  \label{fig:12}
\end{subfigure}\hfil 
\begin{subfigure}{0.17\textwidth}
  \includegraphics[width=0.95\linewidth, frame]{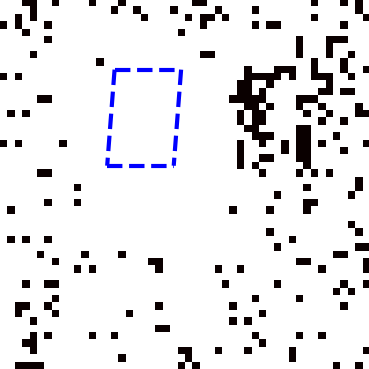}
  \caption{FD + MOG\textsuperscript{2}}
  \label{fig:13}
\end{subfigure}\hfil
\begin{subfigure}{0.17\textwidth}
  \includegraphics[width=0.95\linewidth, frame]{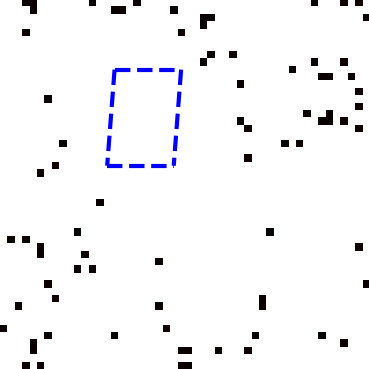}
  \caption{FD + KNN\textsuperscript{2}}
  \label{fig:14}
\end{subfigure}\hfil 
\begin{subfigure}{0.17\textwidth}
  \includegraphics[width=0.95\linewidth, frame]{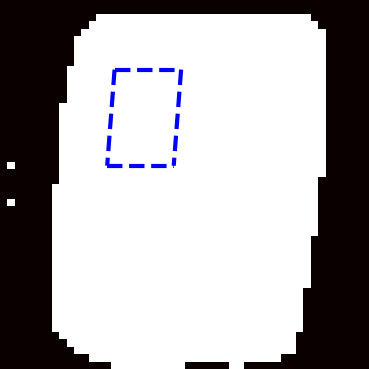}
  \caption{Otsu\textsuperscript{2}}
  \label{fig:15}
\end{subfigure}\hfil 

\medskip

\begin{subfigure}{0.17\textwidth}
  \includegraphics[width=0.95\linewidth, frame]{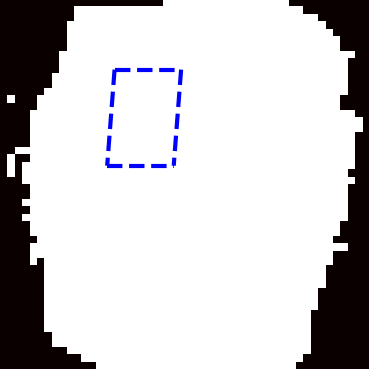}
  \caption{Yen\textsuperscript{2}}
  \label{fig:17}
\end{subfigure}\hfil 
\begin{subfigure}{0.17\textwidth}
  \includegraphics[width=0.95\linewidth, frame]{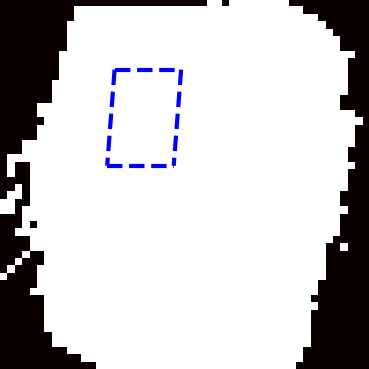}
  \caption{Li\textsuperscript{2}}
  \label{fig:16}
\end{subfigure}\hfil 
\begin{subfigure}{0.17\textwidth}
  \includegraphics[width=0.95\linewidth, frame]{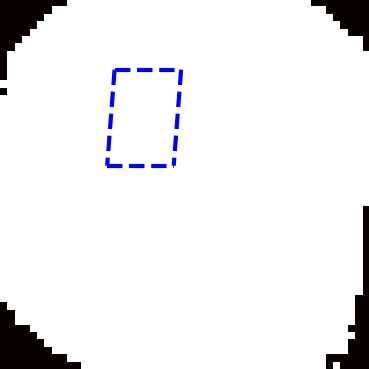}
  \caption{Triangle\textsuperscript{2}}
  \label{fig:18}
\end{subfigure}\hfil 
\begin{subfigure}{0.17\textwidth}
  \includegraphics[width=0.95\linewidth, frame]{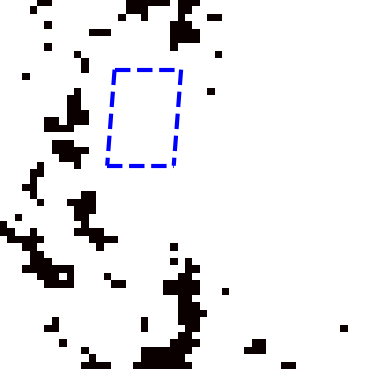}
  \caption{FD + GSOC\textsuperscript{2}}
  \label{fig:17}
\end{subfigure}\hfil 
\begin{subfigure}{0.17\textwidth}
  \includegraphics[width=0.95\linewidth, frame]{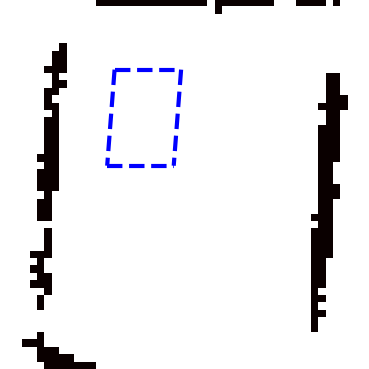}
  \caption{AdaptMean\textsuperscript{2}}
  \label{fig:19}
\end{subfigure}\hfil 

\medskip

\begin{subfigure}{0.17\textwidth}
  \includegraphics[width=0.95\linewidth, frame]{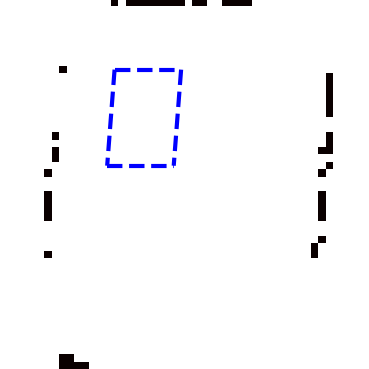}
  \caption{AdaptGauss\textsuperscript{2}}
  \label{fig:20}
\end{subfigure}\hfil 
\begin{subfigure}{0.17\textwidth}
  \includegraphics[width=0.95\linewidth, frame]{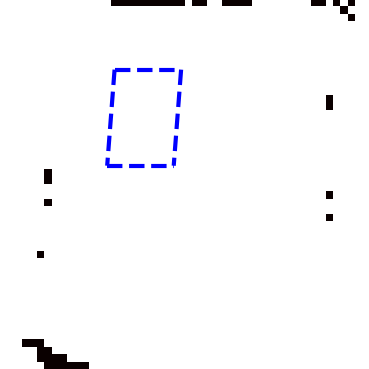}
  \caption{Sauvola\textsuperscript{2}}
  \label{fig:22}
\end{subfigure}\hfil 
\begin{subfigure}{0.17\textwidth}
  \includegraphics[width=0.95\linewidth, frame]{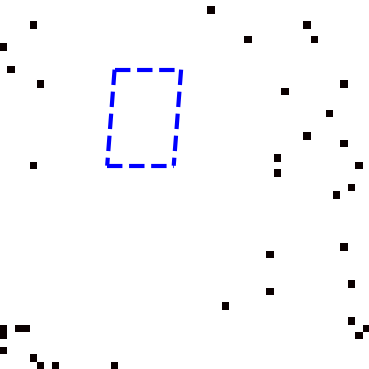}
  \caption{FD\textsuperscript{2}}
  \label{fig:23}
\end{subfigure}\hfil
\begin{subfigure}{0.17\textwidth}
  \includegraphics[width=0.95\linewidth, frame]{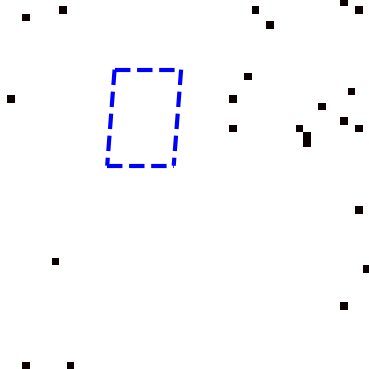}
  \caption{FD + CNT\textsuperscript{2}}
  \label{fig:24}
\end{subfigure}\hfil 
\begin{subfigure}{0.17\textwidth}
  \includegraphics[width=0.95\linewidth, frame]{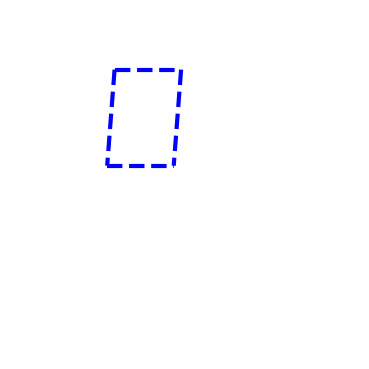}
  \caption{Others\textsuperscript{2}}
  \label{fig:25}
\end{subfigure}\hfil 

\caption{Binary composite images with projected contours for the SLM dataset. Superscript 1 indicates no spatter detection was observed in the foregrounds, superscript 2 indicates that the foregrounds exhibited were impacted by ejected spatter or had general segmentation problems.}
\label{fig:SLM_binary_composite_images}
\end{figure*}

Figure~\ref{fig:SLM_binary_composite_images} shows the binary composite image for each algorithm. The white colors represent nonzero pixel values, while the black colors represent values of zero. The dashed boxes are the projected contours and indicate the area of the upper-left cross-section from Figure~\ref{fig:mask_sub2}. 

For the algorithms of the first row (MOG, MOG2, CNT, GSOC, and Thresh) and the KNN algorithm in the second row of Figure~\ref{fig:SLM_binary_composite_images}, the areas of the nonzero pixels approximately match the areas of the projected contours. The nonzero pixels in the composite images from these algorithms are either inside the projected contours or part of the overflow of nonzero pixels around the contours. Aligned with our manual inspection of the segmented results, these algorithms did not exhibit any artifacts in their foregrounds. Each foreground only featured nonzero pixels that approximately represented the most recent laser track changes on the build surface. Note that KNN has a single isolated nonzero pixel laying outside of the contour. Based on our manual inspection, this detection stems from general segmentation problems of the KNN algorithm due to poor segmentation accuracy (see Section~\ref{Seg_acc}).

The remaining algorithms were either significantly influenced by the spatter or had general segmentation problems. The segmented results from these algorithms exhibited numerous artifacts in each foreground. These artifacts may stem from the ejected spatter or general segmentation problems of the algorithm. As a result, the composite images for most of these algorithms exhibit numerous nonzero pixels outside of the projected contours, some even consisting almost entirely of nonzero pixels. Only the Thresh($\lambda$) and GMG algorithms did not exhibit this characteristic. Instead, these algorithms missed foreground pixels during scanning.

\section{Discussion} \label{section_6}

We found that the combination of frame differencing with common thresholding and background segmentation algorithms performed best for finding the most recent changes of the laser tracks in each infrared image. Especially the FD+Thresh($\lambda$) algorithm, which combines frame differencing with simple thresholding, performed better than any other segmentation algorithm while also maintaining one of the highest speeds for processing each image. Furthermore, we found that all of the popular thresholding algorithms that were recently used in LPBF studies, such as Otsu's method, performed poorly on the segmentation accuracy test. 

The FD+Thresh($\lambda$) algorithm has the advantage that the $\lambda$ parameter is relatively easy to tune compared to the other segmentation algorithms. If the desired width for the detected laser track should be thinner, $\lambda$ can be increased to eliminate more foreground pixels from the border of the laser track; if the desired width should be bigger, then $\lambda$ can be decreased. However, there might be LPBF process setups where the default $\lambda$ does not work well, and calibrating the algorithm is not possible. In this case, the parameterless combination algorithm FD+Yen is a promising alternative that has similar speeds as the FD+Thresh($\lambda$) algorithm and still has better performance than the baseline algorithm Thresh($\lambda$). 

Algorithms like FD+Thresh($\lambda$) can be readily deployed to address several severe process limitations for LPBF processes. First, the FD+Thresh($\lambda$) algorithm can be used to online monitor the most recent laser track changes with respect to different process signatures, such as temperature distribution and geometrical dimensions. Even the entire laser tracks and the heat-affected zone could be monitored with the help of the segmented outputs, as described in Section~\ref{section_2}. Second, the segmented outputs can be used to reduce noise and dimensionality of the data for data-driven methods by eliminating the uninformative portions of the data, such as the backgrounds and the images where the laser is off. This pre-processing of the data can significantly increase the performance of the data-driven methods. Finally, the promise of extracting the informative portions of the images can also be applied to store the images efficiently. Since high-speed infrared cameras produce massive amounts of data, using the segmented outputs to store only the informative portions of the data significantly reduces the memory footprint and addresses current data storage limitations.

The second key finding of this study is that only a few algorithms (MOG, MOG2, CNT, GSOC, Thresh($\lambda$), KNN) successfully excluded spatter from the foregrounds. However, none of these algorithms were among the top-performing algorithms in Section~\ref{Seg_acc}; these six algorithms only had mediocre to poor segmentation accuracy. Nevertheless, the baseline algorithm Thresh($\lambda$) still approximately detected the changes in the laser tracks and may prove useful for processes with strong spatter characteristics, where an approximate detection of the most recent changes of the laser tracks suffices. 

With regards to the qualitative nature of the manual inspection, it is important to note that the manual inspection only sufficed to identify which algorithms do \textit{not} detect spatter in their foregrounds. The manual inspection was not sufficient to differentiate between poor segmentation accuracy and spatter detection for the algorithms with artifacts in the foregrounds. Some of the algorithms that exhibited artifacts in the foregrounds on the SLM dataset may perform differently with non-default parameters. Further investigation is needed to quantify the degree of spatter detection for each algorithm and differentiate between segmentation accuracy and spatter detection. Nevertheless, the current findings may provide important insights into the effect of ejected spatter on the segmentation algorithms. 

Overall, based on the two key findings of this study, none of the algorithms currently possess high segmentation accuracy, high computational speed while also being free of foreground artifacts for processes with strong spatter characteristics. Finding segmentation algorithms that combine these characteristics still requires further investigation. Investigating the degree of spatter detection may also be useful for applications where spatter detection is desired.

\section{Conclusion} \label{section_7}

We studied different segmentation algorithms that separate each infrared image into a foreground and background to address key process limitations for the laser powder bed fusion (LPBF) processes. We present a summary of the results in Table~\ref{table:all_results1} and \ref{table:all_results2}. By evaluating each algorithm based on segmentation accuracy and computational speed, we found that the combination of frame differencing methods with simple thresholding, called FD+Thresh($\lambda$), works best for the LPBF processes. This algorithm had the best segmentation accuracy and also one of the fastest computational speeds. Due to its simple implementation and intuitive parameter, the FD+Thresh($\lambda$) algorithm can be readily applied to online monitor the most recent laser track changes and extract the informative portions of the infrared images for pre-processing and efficient data storage.

We also qualitatively analyzed the spatter detection for each algorithm as the ejected spatter can significantly influence the image segmentation. We found that only a few algorithms did not include spatter in the foregrounds. However, those algorithms possessed mediocre segmentation accuracy at best. Further investigations are needed to find algorithms that combine high segmentation accuracy and computational speed with robustness against spatter detection. 
\section*{Acknowledgments}

This material is based upon work supported by the National Science Foundation under Grant No. 1646522. 

The authors would like to thank Alexander Shkoruta and Dr. Sandipan Mishra at the Rensselaer Polytechnic Institute for providing the Selective Laser Melting data. 

The authors would also like to acknowledge the instrumental research support of the US Department of Defense (DoD) Air Force Research Laboratory (AFRL) Project \# FA8650-17-C-5716 P00003: Laser Additive Manufacturing Pilot Scale (LAMPS) III: Advanced Process Monitoring and Control (PI: Scott Fish)



\bibliographystyle{elsarticle-harv}


\bibliography{refs}

\begin{thebibliography}{57}
\expandafter\ifx\csname natexlab\endcsname\relax\def\natexlab#1{#1}\fi
\providecommand{\url}[1]{\texttt{#1}}
\providecommand{\href}[2]{#2}
\providecommand{\path}[1]{#1}
\providecommand{\DOIprefix}{doi:}
\providecommand{\ArXivprefix}{arXiv:}
\providecommand{\URLprefix}{URL: }
\providecommand{\Pubmedprefix}{pmid:}
\providecommand{\doi}[1]{\href{http://dx.doi.org/#1}{\path{#1}}}
\providecommand{\Pubmed}[1]{\href{pmid:#1}{\path{#1}}}
\providecommand{\bibinfo}[2]{#2}
\ifx\xfnm\relax \def\xfnm[#1]{\unskip,\space#1}\fi
\bibitem[{Baumgartl et~al.(2020)Baumgartl, Tomas, Buettner and Merkel}]{DL_CNN}
\bibinfo{author}{Baumgartl, H.}, \bibinfo{author}{Tomas, J.},
  \bibinfo{author}{Buettner, R.}, \bibinfo{author}{Merkel, M.},
  \bibinfo{year}{2020}.
\newblock \bibinfo{title}{A deep learning-based model for defect detection in
  laser-powder bed fusion using in-situ thermographic monitoring}
  \DOIprefix\doi{10.1007/s40964-019-00108-3}.
\bibitem[{Bouwmans(2014)}]{review_paper_2}
\bibinfo{author}{Bouwmans, T.}, \bibinfo{year}{2014}.
\newblock \bibinfo{title}{Traditional and recent approaches in background
  modeling for foreground detection: An overview}.
\newblock \bibinfo{journal}{Computer Science Review} \bibinfo{volume}{11}.
\newblock \DOIprefix\doi{10.1016/j.cosrev.2014.04.001}.
\bibitem[{Bouwmans et~al.(2014)Bouwmans, Porikli, Höferlin and
  Vacavant}]{bouwman_book}
\bibinfo{author}{Bouwmans, T.}, \bibinfo{author}{Porikli, F.},
  \bibinfo{author}{Höferlin, B.}, \bibinfo{author}{Vacavant, A.},
  \bibinfo{year}{2014}.
\newblock \bibinfo{title}{Handbook on "Background Modeling and Foreground
  Detection for Video Surveillance"}.
\newblock \DOIprefix\doi{10.1201/b17223}.
\bibitem[{Bradski(2000)}]{opencv_library}
\bibinfo{author}{Bradski, G.}, \bibinfo{year}{2000}.
\newblock \bibinfo{title}{{The OpenCV Library}}.
\newblock \bibinfo{journal}{Dr. Dobb's Journal of Software Tools} .
\bibitem[{Chaki et~al.(2014)Chaki, Shaikh and
  Saeed}]{image_binarization_reference}
\bibinfo{author}{Chaki, N.}, \bibinfo{author}{Shaikh, S.},
  \bibinfo{author}{Saeed, K.}, \bibinfo{year}{2014}.
\newblock \bibinfo{title}{A Comprehensive Survey on Image Binarization
  Techniques}.
\newblock pp. \bibinfo{pages}{5--15}.
\newblock \DOIprefix\doi{10.1007/978-81-322-1907-1\_2}.
\bibitem[{Clijsters et~al.(2014)Clijsters, Craeghs, Buls, Kempen and
  Kruth}]{fancy_composite_image}
\bibinfo{author}{Clijsters, S.}, \bibinfo{author}{Craeghs, T.},
  \bibinfo{author}{Buls, S.}, \bibinfo{author}{Kempen, K.},
  \bibinfo{author}{Kruth, J.P.}, \bibinfo{year}{2014}.
\newblock \bibinfo{title}{In situ quality control of the selective laser
  melting process using a high-speed, real-time melt pool monitoring system}.
\newblock \bibinfo{journal}{International Journal of Advanced Manufacturing
  Technology} \bibinfo{volume}{75}, \bibinfo{pages}{1089--1101}.
\newblock \DOIprefix\doi{10.1007/s00170-014-6214-8}.
\bibitem[{Everton et~al.(2016)Everton, Hirsch, Stavroulakis, Leach and
  Clare}]{slm_review}
\bibinfo{author}{Everton, S.}, \bibinfo{author}{Hirsch, M.},
  \bibinfo{author}{Stavroulakis, P.}, \bibinfo{author}{Leach, R.},
  \bibinfo{author}{Clare, A.}, \bibinfo{year}{2016}.
\newblock \bibinfo{title}{Review of in-situ process monitoring and in-situ
  metrology for metal additive manufacturing}.
\newblock \bibinfo{journal}{Materials \& Design} \bibinfo{volume}{95},
  \bibinfo{pages}{431--445}.
\newblock \DOIprefix\doi{10.1016/j.matdes.2016.01.099}.
\bibitem[{Godbehere et~al.(2012)Godbehere, Matsukawa and Goldberg}]{GMG}
\bibinfo{author}{Godbehere, A.}, \bibinfo{author}{Matsukawa, A.},
  \bibinfo{author}{Goldberg, K.}, \bibinfo{year}{2012}.
\newblock \bibinfo{title}{Visual tracking of human visitors under
  variable-lighting conditions for a responsive audio art installation}, pp.
  \bibinfo{pages}{4305--4312}.
\newblock \DOIprefix\doi{10.1109/ACC.2012.6315174}.
\bibitem[{Grasso and Colosimo(2017)}]{SLM_review_insitu_monitoring}
\bibinfo{author}{Grasso, M.}, \bibinfo{author}{Colosimo, B.},
  \bibinfo{year}{2017}.
\newblock \bibinfo{title}{Process defects and in-situ monitoring methods in
  metal powder bed fusion: a review}.
\newblock \bibinfo{journal}{Measurement Science and Technology}
  \bibinfo{volume}{28}, \bibinfo{pages}{1--25}.
\newblock \DOIprefix\doi{10.1088/1361-6501/aa5c4f}.
\bibitem[{Grasso and Colosimo(2019)}]{Zinc_successor_paper}
\bibinfo{author}{Grasso, M.}, \bibinfo{author}{Colosimo, B.},
  \bibinfo{year}{2019}.
\newblock \bibinfo{title}{A statistical learning method for image-based
  monitoring of the plume signature in laser powder bed fusion}.
\newblock \bibinfo{journal}{Robotics and Computer-Integrated Manufacturing}
  \bibinfo{volume}{57}, \bibinfo{pages}{103 -- 115}.
\newblock \DOIprefix\doi{https://doi.org/10.1016/j.rcim.2018.11.007}.
\bibitem[{Grasso et~al.(2018)Grasso, Demir and
  Colosimo}]{insitu_monitoring_SLM_zinc_powder}
\bibinfo{author}{Grasso, M.}, \bibinfo{author}{Demir, A.G.},
  \bibinfo{author}{Colosimo, B.}, \bibinfo{year}{2018}.
\newblock \bibinfo{title}{In situ monitoring of selective laser melting of zinc
  powder via infrared imaging of the process plume}.
\newblock \bibinfo{journal}{Robotics and Computer-Integrated Manufacturing}
  \bibinfo{volume}{49}, \bibinfo{pages}{229--239}.
\newblock \DOIprefix\doi{10.1016/j.rcim.2017.07.001}.
\bibitem[{Guo et~al.(2016)Guo, Xu and Qiang}]{LSBP}
\bibinfo{author}{Guo, L.}, \bibinfo{author}{Xu, D.}, \bibinfo{author}{Qiang,
  Z.}, \bibinfo{year}{2016}.
\newblock \bibinfo{title}{Background subtraction using local svd binary
  pattern}, pp. \bibinfo{pages}{1159--1167}.
\newblock \DOIprefix\doi{10.1109/CVPRW.2016.148}.
\bibitem[{Harris et~al.(2020)Harris, Millman, van~der Walt, Gommers, Virtanen,
  Cournapeau, Wieser, Taylor, Berg, Smith, Kern, Picus, Hoyer, van Kerkwijk,
  Brett, Haldane, Fernández~del Río, Wiebe, Peterson, Gérard-Marchant,
  Sheppard, Reddy, Weckesser, Abbasi, Gohlke and Oliphant}]{numpy}
\bibinfo{author}{Harris, C.R.}, \bibinfo{author}{Millman, K.J.},
  \bibinfo{author}{van~der Walt, S.J.}, \bibinfo{author}{Gommers, R.},
  \bibinfo{author}{Virtanen, P.}, \bibinfo{author}{Cournapeau, D.},
  \bibinfo{author}{Wieser, E.}, \bibinfo{author}{Taylor, J.},
  \bibinfo{author}{Berg, S.}, \bibinfo{author}{Smith, N.J.},
  \bibinfo{author}{Kern, R.}, \bibinfo{author}{Picus, M.},
  \bibinfo{author}{Hoyer, S.}, \bibinfo{author}{van Kerkwijk, M.H.},
  \bibinfo{author}{Brett, M.}, \bibinfo{author}{Haldane, A.},
  \bibinfo{author}{Fernández~del Río, J.}, \bibinfo{author}{Wiebe, M.},
  \bibinfo{author}{Peterson, P.}, \bibinfo{author}{Gérard-Marchant, P.},
  \bibinfo{author}{Sheppard, K.}, \bibinfo{author}{Reddy, T.},
  \bibinfo{author}{Weckesser, W.}, \bibinfo{author}{Abbasi, H.},
  \bibinfo{author}{Gohlke, C.}, \bibinfo{author}{Oliphant, T.E.},
  \bibinfo{year}{2020}.
\newblock \bibinfo{title}{Array programming with {NumPy}}.
\newblock \bibinfo{journal}{Nature} \bibinfo{volume}{585},
  \bibinfo{pages}{357–362}.
\newblock \DOIprefix\doi{10.1038/s41586-020-2649-2}.
\bibitem[{{Jui-Cheng Yen} et~al.(1995){Jui-Cheng Yen}, {Fu-Juay Chang} and
  {Shyang Chang}}]{Yen}
\bibinfo{author}{{Jui-Cheng Yen}}, \bibinfo{author}{{Fu-Juay Chang}},
  \bibinfo{author}{{Shyang Chang}}, \bibinfo{year}{1995}.
\newblock \bibinfo{title}{A new criterion for automatic multilevel
  thresholding}.
\newblock \bibinfo{journal}{IEEE Transactions on Image Processing}
  \bibinfo{volume}{4}, \bibinfo{pages}{370--378}.
\bibitem[{Kaewtrakulpong and Bowden(2002)}]{MOG}
\bibinfo{author}{Kaewtrakulpong, P.}, \bibinfo{author}{Bowden, R.},
  \bibinfo{year}{2002}.
\newblock \bibinfo{title}{An improved adaptive background mixture model for
  realtime tracking with shadow detection}.
\newblock \bibinfo{journal}{Proceedings of 2nd European Workshop on Advanced
  Video-Based Surveillance Systems; September 4, 2001; London, U.K}
  \DOIprefix\doi{10.1007/978-1-4615-0913-4\_11}.
\bibitem[{Kotsiantis et~al.(2006)Kotsiantis, Kanellopoulos and
  Pintelas}]{importance_preprocessing_2}
\bibinfo{author}{Kotsiantis, S.}, \bibinfo{author}{Kanellopoulos, D.},
  \bibinfo{author}{Pintelas, P.}, \bibinfo{year}{2006}.
\newblock \bibinfo{title}{Data preprocessing for supervised learning}.
\newblock \bibinfo{journal}{International Journal of Computer Science}
  \bibinfo{volume}{1}, \bibinfo{pages}{111--117}.
\bibitem[{Krauss et~al.(2012)Krauss, Eschey and Z{\"a}h}]{slm_NIST_cited_this}
\bibinfo{author}{Krauss, H.}, \bibinfo{author}{Eschey, C.},
  \bibinfo{author}{Z{\"a}h, M.F.}, \bibinfo{year}{2012}.
\newblock \bibinfo{title}{Thermography for monitoring the selective laser
  melting process}.
\bibitem[{Krauss et~al.(2014)Krauss, Zeugner and
  Zaeh}]{slm_NIST_cited_this_successor}
\bibinfo{author}{Krauss, H.}, \bibinfo{author}{Zeugner, T.},
  \bibinfo{author}{Zaeh, M.F.}, \bibinfo{year}{2014}.
\newblock \bibinfo{title}{Layerwise monitoring of the selective laser melting
  process by thermography}.
\newblock \bibinfo{journal}{Physics Procedia} \bibinfo{volume}{56},
  \bibinfo{pages}{64 -- 71}.
\newblock \DOIprefix\doi{https://doi.org/10.1016/j.phpro.2014.08.097}.
  \bibinfo{note}{8th International Conference on Laser Assisted Net Shape
  Engineering LANE 2014}.
\bibitem[{Kruth et~al.(2005)Kruth, Vandenbroucke, Vaerenbergh and
  Mercelis}]{3_SLS_SLM_application}
\bibinfo{author}{Kruth, J.P.}, \bibinfo{author}{Vandenbroucke, B.},
  \bibinfo{author}{Vaerenbergh, J.}, \bibinfo{author}{Mercelis, P.},
  \bibinfo{year}{2005}.
\newblock \bibinfo{title}{Benchmarking of different sls/slm processes as rapid
  manufacturing techniques}.
\newblock \bibinfo{journal}{IEEE Electron Device Letters - IEEE ELECTRON DEV
  LETT} \bibinfo{volume}{10}.
\bibitem[{Li and Tam(1998)}]{Li}
\bibinfo{author}{Li, C.}, \bibinfo{author}{Tam, P.}, \bibinfo{year}{1998}.
\newblock \bibinfo{title}{An iterative algorithm for minimum cross entropy
  thresholding}.
\newblock \bibinfo{journal}{Pattern Recognition Letters} \bibinfo{volume}{19},
  \bibinfo{pages}{771 -- 776}.
\newblock \DOIprefix\doi{https://doi.org/10.1016/S0167-8655(98)00057-9}.
\bibitem[{Lo et~al.(2019)Lo, Liu and Tran}]{optimized_hatch_space}
\bibinfo{author}{Lo, Y.L.}, \bibinfo{author}{Liu, B.Y.}, \bibinfo{author}{Tran,
  H.C.}, \bibinfo{year}{2019}.
\newblock \bibinfo{title}{Optimized hatch space selection in double-scanning
  track selective laser melting process}.
\newblock \bibinfo{journal}{The International Journal of Advanced Manufacturing
  Technology} \bibinfo{volume}{105}.
\newblock \DOIprefix\doi{10.1007/s00170-019-04456-w}.
\bibitem[{Louvis et~al.(2011)Louvis, Fox and Sutcliffe}]{study_of_aluminum}
\bibinfo{author}{Louvis, E.}, \bibinfo{author}{Fox, P.},
  \bibinfo{author}{Sutcliffe, C.J.}, \bibinfo{year}{2011}.
\newblock \bibinfo{title}{Selective laser melting of aluminium components}.
\newblock \bibinfo{journal}{Journal of Materials Processing Technology}
  \bibinfo{volume}{211}, \bibinfo{pages}{275 -- 284}.
\newblock \DOIprefix\doi{https://doi.org/10.1016/j.jmatprotec.2010.09.019}.
\bibitem[{Lu et~al.(2004)Lu, Mausel, Brondízio and
  Moran}]{change_detection_techniques}
\bibinfo{author}{Lu, D.}, \bibinfo{author}{Mausel, P.},
  \bibinfo{author}{Brondízio, E.}, \bibinfo{author}{Moran, E.},
  \bibinfo{year}{2004}.
\newblock \bibinfo{title}{Change detection techniques. int j remote sens}.
\newblock \bibinfo{journal}{International Journal of Remote Sensing}
  \bibinfo{volume}{25}, \bibinfo{pages}{2365--2401}.
\newblock \DOIprefix\doi{10.1080/0143116031000139863}.
\bibitem[{Mazzoleni et~al.(2019)Mazzoleni, Demir, Caprio, Pacher and
  Previtali}]{overview_of_off_axis_approaches}
\bibinfo{author}{Mazzoleni, L.}, \bibinfo{author}{Demir, A.G.},
  \bibinfo{author}{Caprio, L.}, \bibinfo{author}{Pacher, M.},
  \bibinfo{author}{Previtali, B.}, \bibinfo{year}{2019}.
\newblock \bibinfo{title}{Real-time observation of melt pool in selective laser
  melting: Spatial, temporal and wavelength resolution criteria}.
\newblock \bibinfo{journal}{IEEE Transactions on Instrumentation and
  Measurement} \bibinfo{volume}{PP}, \bibinfo{pages}{1--1}.
\newblock \DOIprefix\doi{10.1109/TIM.2019.2912236}.
\bibitem[{Nadammal et~al.(2017)Nadammal, Cabeza, Mishurova, Thiede, Kromm,
  Seyfert, Farahbod, Haberland, Schneider, Portella and
  Bruno}]{effect_of_hatch_length}
\bibinfo{author}{Nadammal, N.}, \bibinfo{author}{Cabeza, S.},
  \bibinfo{author}{Mishurova, T.}, \bibinfo{author}{Thiede, T.},
  \bibinfo{author}{Kromm, A.}, \bibinfo{author}{Seyfert, C.},
  \bibinfo{author}{Farahbod, L.}, \bibinfo{author}{Haberland, C.},
  \bibinfo{author}{Schneider, J.A.}, \bibinfo{author}{Portella, P.D.},
  \bibinfo{author}{Bruno, G.}, \bibinfo{year}{2017}.
\newblock \bibinfo{title}{Effect of hatch length on the development of
  microstructure, texture and residual stresses in selective laser melted
  superalloy inconel 718}.
\newblock \bibinfo{journal}{Materials \& Design} \bibinfo{volume}{134},
  \bibinfo{pages}{139 -- 150}.
\newblock \DOIprefix\doi{https://doi.org/10.1016/j.matdes.2017.08.049}.
\bibitem[{Nawi et~al.(2013)Nawi, Atomi and Rehman}]{importance_preprocessing}
\bibinfo{author}{Nawi, N.M.}, \bibinfo{author}{Atomi, W.H.},
  \bibinfo{author}{Rehman, M.}, \bibinfo{year}{2013}.
\newblock \bibinfo{title}{The effect of data pre-processing on optimized
  training of artificial neural networks}.
\newblock \bibinfo{journal}{Procedia Technology} \bibinfo{volume}{11},
  \bibinfo{pages}{32 -- 39}.
\newblock \DOIprefix\doi{https://doi.org/10.1016/j.protcy.2013.12.159}.
  \bibinfo{note}{4th International Conference on Electrical Engineering and
  Informatics, ICEEI 2013}.
\bibitem[{{Otsu}(1979)}]{Otsu}
\bibinfo{author}{{Otsu}, N.}, \bibinfo{year}{1979}.
\newblock \bibinfo{title}{A threshold selection method from gray-level
  histograms}.
\newblock \bibinfo{journal}{IEEE Transactions on Systems, Man, and Cybernetics}
  \bibinfo{volume}{9}, \bibinfo{pages}{62--66}.
\bibitem[{Pereira et~al.(2019)Pereira, Kennedy and Potgieter}]{1_trad_manufact}
\bibinfo{author}{Pereira, T.}, \bibinfo{author}{Kennedy, J.V.},
  \bibinfo{author}{Potgieter, J.}, \bibinfo{year}{2019}.
\newblock \bibinfo{title}{A comparison of traditional manufacturing vs additive
  manufacturing, the best method for the job}.
\newblock \bibinfo{journal}{Procedia Manufacturing} \bibinfo{volume}{30},
  \bibinfo{pages}{11 -- 18}.
\newblock \DOIprefix\doi{10.1016/j.promfg.2019.02.003}. \bibinfo{note}{{Digital
  Manufacturing Transforming Industry Towards Sustainable Growth}}.
\bibitem[{Pulli et~al.(2012)Pulli, Baksheev, Kornyakov and
  Eruhimov}]{openCV_real_time}
\bibinfo{author}{Pulli, K.}, \bibinfo{author}{Baksheev, A.},
  \bibinfo{author}{Kornyakov, K.}, \bibinfo{author}{Eruhimov, V.},
  \bibinfo{year}{2012}.
\newblock \bibinfo{title}{Real-time computer vision with opencv}.
\newblock \bibinfo{journal}{Communications of the ACM} \bibinfo{volume}{55},
  \bibinfo{pages}{61--69}.
\newblock \DOIprefix\doi{10.1145/2184319.2184337}.
\bibitem[{Radke et~al.(2005)Radke, Andra, Al-Kofahi and
  Roysam}]{change_detection_review}
\bibinfo{author}{Radke, R.}, \bibinfo{author}{Andra, S.},
  \bibinfo{author}{Al-Kofahi, O.}, \bibinfo{author}{Roysam, B.},
  \bibinfo{year}{2005}.
\newblock \bibinfo{title}{Image change detection algorithms: A systematic
  survey}.
\newblock \bibinfo{journal}{IEEE transactions on image processing : a
  publication of the IEEE Signal Processing Society} \bibinfo{volume}{14},
  \bibinfo{pages}{294--307}.
\newblock \DOIprefix\doi{10.1109/TIP.2004.838698}.
\bibitem[{Ridler(1978)}]{isodata}
\bibinfo{author}{Ridler, T.W.}, \bibinfo{year}{1978}.
\newblock \bibinfo{title}{Picture thresholding using an iterative selection
  method.}
\newblock \bibinfo{journal}{IEEE Transactions on Systems, Man, and Cybernetics}
  \bibinfo{volume}{8}, \bibinfo{pages}{630--632}.
\bibitem[{Robinson et~al.(2018)Robinson, Ashton, Jones, Fox and
  Sutcliffe}]{effect_hatch_angle_on_parts}
\bibinfo{author}{Robinson, J.}, \bibinfo{author}{Ashton, I.},
  \bibinfo{author}{Jones, E.}, \bibinfo{author}{Fox, P.},
  \bibinfo{author}{Sutcliffe, C.}, \bibinfo{year}{2018}.
\newblock \bibinfo{title}{The effect of hatch angle rotation on parts
  manufactured using selective laser melting}.
\newblock \bibinfo{journal}{Rapid Prototyping Journal} \bibinfo{volume}{25}.
\newblock \DOIprefix\doi{10.1108/RPJ-06-2017-0111}.
\bibitem[{Rosin and Ellis(1995)}]{image_difference_threshold}
\bibinfo{author}{Rosin, P.}, \bibinfo{author}{Ellis, T.}, \bibinfo{year}{1995}.
\newblock \bibinfo{title}{Image difference threshold strategies and shadow
  detection} \bibinfo{volume}{1}.
\newblock \DOIprefix\doi{10.5244/C.9.35}.
\bibitem[{Rosin and Ioannidis(2003)}]{change_detection_thresholding}
\bibinfo{author}{Rosin, P.}, \bibinfo{author}{Ioannidis, E.},
  \bibinfo{year}{2003}.
\newblock \bibinfo{title}{Evaluation of global image thresholding for change
  detection}.
\newblock \bibinfo{journal}{Pattern Recognition Letters} \bibinfo{volume}{24},
  \bibinfo{pages}{2345--2356}.
\newblock \DOIprefix\doi{10.1016/S0167-8655(03)00060-6}.
\bibitem[{Sauvola and Pietikäinen(2000)}]{sauvola}
\bibinfo{author}{Sauvola, J.}, \bibinfo{author}{Pietikäinen, M.},
  \bibinfo{year}{2000}.
\newblock \bibinfo{title}{Adaptive document image binarization}.
\newblock \bibinfo{journal}{Pattern Recognition} \bibinfo{volume}{33},
  \bibinfo{pages}{225 -- 236}.
\newblock \DOIprefix\doi{https://doi.org/10.1016/S0031-3203(99)00055-2}.
\bibitem[{Sezgin and Sankur(2004)}]{image_binarization_survey}
\bibinfo{author}{Sezgin, M.}, \bibinfo{author}{Sankur, B.},
  \bibinfo{year}{2004}.
\newblock \bibinfo{title}{Survey over image thresholding techniques and
  quantitative performance evaluation}.
\newblock \bibinfo{journal}{Journal of Electronic Imaging}
  \bibinfo{volume}{13}, \bibinfo{pages}{146--168}.
\newblock \DOIprefix\doi{10.1117/1.1631315}.
\bibitem[{Shi et~al.(2017)Shi, Ma, Liu and
  Wu}]{effect_process_params_on_track_2}
\bibinfo{author}{Shi, X.}, \bibinfo{author}{Ma, S.}, \bibinfo{author}{Liu, C.},
  \bibinfo{author}{Wu, Q.}, \bibinfo{year}{2017}.
\newblock \bibinfo{title}{Parameter optimization for ti-47al-2cr-2nb in
  selective laser melting based on geometric characteristics of single scan
  tracks}.
\newblock \bibinfo{journal}{Optics \& Laser Technology} \bibinfo{volume}{90},
  \bibinfo{pages}{71 -- 79}.
\newblock \DOIprefix\doi{https://doi.org/10.1016/j.optlastec.2016.11.002}.
\bibitem[{Sobral(2013)}]{openCV_established_MOG_MOG2}
\bibinfo{author}{Sobral, A.}, \bibinfo{year}{2013}.
\newblock \bibinfo{title}{Bgslibrary: An opencv c++ background subtraction
  library}.
\newblock \DOIprefix\doi{10.13140/2.1.1740.7044}.
\bibitem[{Sobral and Vacavant(2014)}]{comparison_most_cited}
\bibinfo{author}{Sobral, A.}, \bibinfo{author}{Vacavant, A.},
  \bibinfo{year}{2014}.
\newblock \bibinfo{title}{A comprehensive review of background subtraction
  algorithms evaluated with synthetic and real videos}.
\newblock \bibinfo{journal}{Computer Vision and Image Understanding}
  \bibinfo{volume}{122}, \bibinfo{pages}{4--21}.
\newblock \DOIprefix\doi{10.1016/j.cviu.2013.12.005}.
\bibitem[{Song et~al.(2015)Song, Zhao, Li, Han, Wei, Wen, Liu and
  Shi}]{2_slm_advantage}
\bibinfo{author}{Song, B.}, \bibinfo{author}{Zhao, X.}, \bibinfo{author}{Li,
  S.}, \bibinfo{author}{Han, C.}, \bibinfo{author}{Wei, Q.},
  \bibinfo{author}{Wen, S.}, \bibinfo{author}{Liu, J.}, \bibinfo{author}{Shi,
  Y.}, \bibinfo{year}{2015}.
\newblock \bibinfo{title}{Differences in microstructure and properties between
  selective laser melting and traditional manufacturing for fabrication of
  metal parts: A review}.
\newblock \bibinfo{journal}{Frontiers of Mechanical Engineering}
  \bibinfo{volume}{10}, \bibinfo{pages}{111--125}.
\newblock \DOIprefix\doi{10.1007/s11465-015-0341-2}.
\bibitem[{Spears and Gold(2016)}]{data_storage_problem}
\bibinfo{author}{Spears, T.}, \bibinfo{author}{Gold, S.}, \bibinfo{year}{2016}.
\newblock \bibinfo{title}{In-process sensing in selective laser melting (slm)
  additive manufacturing}.
\newblock \bibinfo{journal}{Integrating Materials and Manufacturing Innovation}
  \bibinfo{volume}{5}.
\newblock \DOIprefix\doi{10.1186/s40192-016-0045-4}.
\bibitem[{Tapia and Elwany(2014)}]{6_am_review_control}
\bibinfo{author}{Tapia, G.}, \bibinfo{author}{Elwany, A.},
  \bibinfo{year}{2014}.
\newblock \bibinfo{title}{{A Review on Process Monitoring and Control in
  Metal-Based Additive Manufacturing}}.
\newblock \bibinfo{journal}{Journal of Manufacturing Science and Engineering}
  \bibinfo{volume}{136}.
\newblock \DOIprefix\doi{10.1115/1.4028540}. \bibinfo{note}{060801}.
\bibitem[{Tofail et~al.(2018)Tofail, Koumoulos, Bandyopadhyay, Bose,
  O’Donoghue and Charitidis}]{5_am_review_market}
\bibinfo{author}{Tofail, S.A.}, \bibinfo{author}{Koumoulos, E.P.},
  \bibinfo{author}{Bandyopadhyay, A.}, \bibinfo{author}{Bose, S.},
  \bibinfo{author}{O’Donoghue, L.}, \bibinfo{author}{Charitidis, C.},
  \bibinfo{year}{2018}.
\newblock \bibinfo{title}{Additive manufacturing: scientific and technological
  challenges, market uptake and opportunities}.
\newblock \bibinfo{journal}{Materials Today} \bibinfo{volume}{21},
  \bibinfo{pages}{22 -- 37}.
\newblock \DOIprefix\doi{10.1016/j.mattod.2017.07.001}.
\bibitem[{Trnovszky et~al.(2017)Trnovszky, Sykora and
  Hudec}]{MOG_MOG2_KNN_reference}
\bibinfo{author}{Trnovszky, T.}, \bibinfo{author}{Sykora, P.},
  \bibinfo{author}{Hudec, R.}, \bibinfo{year}{2017}.
\newblock \bibinfo{title}{Comparison of background subtraction methods on near
  infra-red spectrum video sequences}.
\newblock \bibinfo{journal}{Procedia Engineering} \bibinfo{volume}{192},
  \bibinfo{pages}{887--892}.
\newblock \DOIprefix\doi{10.1016/j.proeng.2017.06.153}.
\bibitem[{Uriondo et~al.(2015)Uriondo, Esperon~Miguez and
  Perinpanayagam}]{4_SLS_SLM_appl_2}
\bibinfo{author}{Uriondo, A.}, \bibinfo{author}{Esperon~Miguez, M.},
  \bibinfo{author}{Perinpanayagam, S.}, \bibinfo{year}{2015}.
\newblock \bibinfo{title}{The present and future of additive manufacturing in
  the aerospace sector: A review of important aspects}.
\newblock \bibinfo{journal}{Proceedings of the Institution of Mechanical
  Engineers, Part G: Journal of Aerospace Engineering} \bibinfo{volume}{229}.
\newblock \DOIprefix\doi{10.1177/0954410014568797}.
\bibitem[{van~der Walt et~al.(2014)van~der Walt, {S}ch\"onberger,
  {Nunez-Iglesias}, {B}oulogne, {W}arner, {Y}ager, {G}ouillart, {Y}u and the
  scikit-image contributors}]{skimage}
\bibinfo{author}{van~der Walt, S.}, \bibinfo{author}{{S}ch\"onberger, J.L.},
  \bibinfo{author}{{Nunez-Iglesias}, J.}, \bibinfo{author}{{B}oulogne, F.},
  \bibinfo{author}{{W}arner, J.D.}, \bibinfo{author}{{Y}ager, N.},
  \bibinfo{author}{{G}ouillart, E.}, \bibinfo{author}{{Y}u, T.},
  \bibinfo{author}{the scikit-image contributors}, \bibinfo{year}{2014}.
\newblock \bibinfo{title}{scikit-image: image processing in {P}ython}.
\newblock \bibinfo{journal}{PeerJ} \bibinfo{volume}{2}, \bibinfo{pages}{e453}.
\newblock \DOIprefix\doi{10.7717/peerj.453}.
\bibitem[{Williams et~al.(2019)Williams, Ronneberg, Pham, Davies, Hooper,
  Piglione and Jones}]{otsu_plus_watershed_monitoring}
\bibinfo{author}{Williams, R.}, \bibinfo{author}{Ronneberg, T.},
  \bibinfo{author}{Pham, M.S.}, \bibinfo{author}{Davies, C.},
  \bibinfo{author}{Hooper, P.}, \bibinfo{author}{Piglione, A.},
  \bibinfo{author}{Jones, C.}, \bibinfo{year}{2019}.
\newblock \bibinfo{title}{In situ thermography for laser powder bed fusion:
  Effects of layer temperature on porosity, microstructure and mechanical
  properties} \bibinfo{volume}{30}.
\newblock \DOIprefix\doi{10.1016/j.addma.2019.100880}.
\bibitem[{Xu et~al.(2019)Xu, Nettekoven, Julius and Topcu}]{Zhe}
\bibinfo{author}{Xu, Z.}, \bibinfo{author}{Nettekoven, A.},
  \bibinfo{author}{Julius, A.}, \bibinfo{author}{Topcu, U.},
  \bibinfo{year}{2019}.
\newblock \bibinfo{title}{Graph temporal logic inference for classification and
  identification}, pp. \bibinfo{pages}{4761--4768}.
\newblock \DOIprefix\doi{10.1109/CDC40024.2019.9029181}.
\bibitem[{Yadav et~al.(2020)Yadav, Rigo, Arvieu, Guen and
  Lacoste}]{SLM_ML_review}
\bibinfo{author}{Yadav, P.}, \bibinfo{author}{Rigo, O.},
  \bibinfo{author}{Arvieu, C.}, \bibinfo{author}{Guen, E.},
  \bibinfo{author}{Lacoste, E.}, \bibinfo{year}{2020}.
\newblock \bibinfo{title}{In situ monitoring systems of the slm process: On the
  need to develop machine learning models for data processing}.
\newblock \bibinfo{journal}{Crystals} \bibinfo{volume}{10},
  \bibinfo{pages}{524}.
\newblock \DOIprefix\doi{10.3390/cryst10060524}.
\bibitem[{Yadroitsev et~al.(2010)Yadroitsev, Gusarov, Yadroitsava and
  Smurov}]{effect_on_single_track_stability_zones}
\bibinfo{author}{Yadroitsev, I.}, \bibinfo{author}{Gusarov, A.},
  \bibinfo{author}{Yadroitsava, I.}, \bibinfo{author}{Smurov, I.},
  \bibinfo{year}{2010}.
\newblock \bibinfo{title}{Single track formation in selective laser melting of
  metal powders}.
\newblock \bibinfo{journal}{Journal of Materials Processing Technology}
  \bibinfo{volume}{210}, \bibinfo{pages}{1624 -- 1631}.
\newblock \DOIprefix\doi{https://doi.org/10.1016/j.jmatprotec.2010.05.010}.
\bibitem[{Yadroitsev et~al.(2012)Yadroitsev, Yadroitsava, Bertrand and
  Smurov}]{effect_process_params_on_track}
\bibinfo{author}{Yadroitsev, I.}, \bibinfo{author}{Yadroitsava, I.},
  \bibinfo{author}{Bertrand, P.}, \bibinfo{author}{Smurov, I.},
  \bibinfo{year}{2012}.
\newblock \bibinfo{title}{Factor analysis of selective laser melting process
  parameters and geometrical characteristics of synthesized single tracks}.
\newblock \bibinfo{journal}{Rapid Prototyping Journal} \bibinfo{volume}{18},
  \bibinfo{pages}{201--208}.
\newblock \DOIprefix\doi{10.1108/13552541211218117}.
\bibitem[{Yao et~al.(2017)Yao, Lei, Zhong, Jiang and Jia}]{comparison}
\bibinfo{author}{Yao, G.}, \bibinfo{author}{Lei, T.}, \bibinfo{author}{Zhong,
  J.}, \bibinfo{author}{Jiang, P.}, \bibinfo{author}{Jia, W.},
  \bibinfo{year}{2017}.
\newblock \bibinfo{title}{Comparative evaluation of background subtraction
  algorithms in remote scene videos captured by mwir sensors}.
\newblock \bibinfo{journal}{Sensors} \bibinfo{volume}{17},
  \bibinfo{pages}{1945}.
\newblock \DOIprefix\doi{10.3390/s17091945}.
\bibitem[{Ye et~al.(2019)Ye, Zhu, Fuh, Zhang and Soon}]{segmentation_4}
\bibinfo{author}{Ye, D.}, \bibinfo{author}{Zhu, K.}, \bibinfo{author}{Fuh,
  J.Y.H.}, \bibinfo{author}{Zhang, Y.}, \bibinfo{author}{Soon, H.G.},
  \bibinfo{year}{2019}.
\newblock \bibinfo{title}{The investigation of plume and spatter signatures on
  melted states in selective laser melting}.
\newblock \bibinfo{journal}{Optics \& Laser Technology} \bibinfo{volume}{111},
  \bibinfo{pages}{395 -- 406}.
\newblock \DOIprefix\doi{https://doi.org/10.1016/j.optlastec.2018.10.019}.
\bibitem[{Zack et~al.(1977)Zack, Rogers and Latt}]{Triangle}
\bibinfo{author}{Zack, G.W.}, \bibinfo{author}{Rogers, W.E.},
  \bibinfo{author}{Latt, S.}, \bibinfo{year}{1977}.
\newblock \bibinfo{title}{Automatic measurement of sister chromatid exchange
  frequency.}
\newblock \bibinfo{journal}{The journal of histochemistry and cytochemistry :
  official journal of the Histochemistry Society} \bibinfo{volume}{25},
  \bibinfo{pages}{741 -- 753}.
\bibitem[{Zhang et~al.(2018)Zhang, Fuh, Dongsen and Hong}]{segmentation_2}
\bibinfo{author}{Zhang, Y.}, \bibinfo{author}{Fuh, J.},
  \bibinfo{author}{Dongsen, Y.}, \bibinfo{author}{Hong, G.S.},
  \bibinfo{year}{2018}.
\newblock \bibinfo{title}{In-situ monitoring of laser-based pbf via off-axis
  vision and image processing approaches}.
\newblock \bibinfo{journal}{Additive Manufacturing} \bibinfo{volume}{25}.
\newblock \DOIprefix\doi{10.1016/j.addma.2018.10.020}.
\bibitem[{Zivkovic(2004)}]{MOG2}
\bibinfo{author}{Zivkovic, Z.}, \bibinfo{year}{2004}.
\newblock \bibinfo{title}{Improved adaptive gaussian mixture model for
  background subtraction}, pp. \bibinfo{pages}{28 -- 31 Vol.2}.
\newblock \DOIprefix\doi{10.1109/ICPR.2004.1333992}.
\bibitem[{Zivkovic and {van der Heijden}(2006)}]{KNN}
\bibinfo{author}{Zivkovic, Z.}, \bibinfo{author}{{van der Heijden}, F.},
  \bibinfo{year}{2006}.
\newblock \bibinfo{title}{Efficient adaptive density estimation per image pixel
  for the task of background subtraction}.
\newblock \bibinfo{journal}{Pattern Recognition Letters} \bibinfo{volume}{27},
  \bibinfo{pages}{773 -- 780}.
\newblock \DOIprefix\doi{https://doi.org/10.1016/j.patrec.2005.11.005}.

\end{thebibliography}

\clearpage
\appendix

\section{Tables}\label{appendix_tables}
\begin{table}[thb]
    \centering
    \small
    \caption{Parameters for the segmentation algorithms.}
    \captionsetup{justification=centering}
    \begin{tabular}{lll}
        \toprule
        Algorithms & Type & Parameters \\
        \midrule
        
        \multirow{2}{*}{Thresh($\lambda$)}               & calibrated            & $\lambda$: 377.60\\
        {}                                  & default               & $\lambda$: 295 \\
        \addlinespace[5pt]   

        \multirow{2}{*}{FD+Thresh($\lambda$)}        & calibrated            & $\lambda$: 4.44 \\
        {}                                  & default               & $\lambda$: 3 \\
        \addlinespace[5pt]
        
        \multirow{2}{*}{SubMax($\delta$)}        & calibrated            & $\delta$: 125.81 \\
        {}                                  & default               & $\delta$: 20 \\
        \addlinespace[5pt]
        
        \multirow{2}{*}{GSOC}               & calibrated            & 'hitsThresh': 5, 'nSamples': 506, 'propRate': 0.76, 'replaceRate': 0.01\\
        {}                                  & default               & OpenCV default \\
        \addlinespace[5pt]  
        
        \multirow{2}{*}{FD+GSOC}               & calibrated            & 'hitsThresh': 2, 'nSamples': 596, 'propRate': 0.72, 'replaceRate': 0.058 \\
        {}                                  & default               & OpenCV default \\
        \addlinespace[5pt]

        \multirow{2}{*}{CNT}                & calibrated            & 'maxStability': 6, 'minStability': 0 \\
        {}                                  & default               & OpenCV default \\
        \addlinespace[5pt] 
        
        \multirow{2}{*}{FD+CNT}                & calibrated            & 'maxStability': 6, 'minStability': 0 \\
        {}                                  & default               & 0.03 \\
        \addlinespace[5pt]         
        
        \multirow{2}{*}{MOG}                & calibrated            & 'backRatio': 0.47, 'history': 14, 'nmixtures': 290 \\
        {}                                  & default               & OpenCV default \\
        \addlinespace[5pt]
        
        \multirow{2}{*}{FD+MOG}                & calibrated            & 'backRatio': 0.77, 'history': 29, 'nmixtures': 327 \\
        {}                                  & default               & OpenCV default \\
        \addlinespace[5pt]       
        
        \multirow{2}{*}{GMG}                & calibrated            & 'thresh': 0.81\\
        {}                                  & default               & 'thresh': 0.8 \\
        \addlinespace[5pt]    
        
        \multirow{2}{*}{FD+GMG}                & calibrated            & 'thresh': 0.03 \\
        {}                                  & default               & 'thresh': 0.8 \\
        \addlinespace[5pt]

        \multirow{2}{*}{MOG2}               & calibrated            & 'history': 67, 'thresh': 10.10\\
        {}                                  & default               & OpenCV default \\
        \addlinespace[5pt] 
        
        \multirow{2}{*}{FD+MOG2}                & calibrated            & 'history': 27, 'thresh': 1.10 \\
        {}                                  & default               & OpenCV default \\
        \addlinespace[5pt]
        
        \multirow{2}{*}{LSBP}                 & calibrated            & 'radius': 200, 'samples': 266 \\
        {}                                  & default               & OpenCV default \\
        \addlinespace[5pt]
        
        \multirow{2}{*}{FD+LSBP}                 & calibrated            & 'radius': 25, 'samples': 700 \\
        {}                                  & default               & OpenCV default \\
        \addlinespace[5pt]  
        
        \multirow{2}{*}{KNN}                & calibrated            & 'history': 68, 'thresh': 298.81 \\
        {}                                  & default               & OpenCV default \\
        \addlinespace[5pt]
        
        \multirow{2}{*}{FD+KNN}                & calibrated            & 'history': 2, 'thresh': 4.40 \\
        {}                                  & default               & OpenCV default \\
        \addlinespace[5pt]           

        \multirow{2}{*}{AdaptMean}                & calibrated            & 'C': 0, 'box': 451 \\
        {}                                  & default               & OpenCV default \\
        \addlinespace[5pt]    
        
        \multirow{2}{*}{FD+AdaptMean}                & calibrated            &  'C': 66, 'box': 263 \\
        {}                                  & default               & OpenCV default \\
        \addlinespace[5pt]  
        
        \multirow{2}{*}{AdaptGauss}                & calibrated            & 'C': 2, 'box': 387 \\
        {}                                  & default               & OpenCV default \\
        \addlinespace[5pt] 
        
        \multirow{2}{*}{FD+AdaptGauss}                & calibrated            & 'C': 78, 'box': 7 \\
        {}                                  & default               & OpenCV default \\
        \addlinespace[5pt]
        
        \multirow{2}{*}{Sauvola}                & calibrated            & 'box': 79, 'k': 0.41\\
        {}                                  & default               & scikit-image default \\
        \addlinespace[5pt] 

        \multirow{2}{*}{FD+Sauvola}                & calibrated            & 'box': 37, 'k': 0.23 \\
        {}                                  & default               & scikit-image default \\
        \addlinespace[5pt]

        \bottomrule
    \end{tabular}
    \label{table:parameters}
\end{table}

\begin{table}
    \centering
    \caption{Performance summary of all algorithms (Part 1).}
    \captionsetup{justification=centering}
    \begin{tabular}{llccccc}
        \toprule
        \multicolumn{2}{c}{\multirow{2}{*}[-1em]{Algorithms}} & \multicolumn{3}{c}{Segmentation accuracy} & Computation time & {Spatter robust?} \\
        \addlinespace[10pt]
        \cmidrule(lr){3-5} \cmidrule(lr){6-6} \cmidrule(lr){7-7}\\
        {} & {} & Precision & Recall & F\textsubscript{1}-score  & [\si{ms}] \\
        \midrule
        
\multirow{2}{*}{Thresh($\lambda$)} & default        & 0.09 & 0.98 & 0.17 & \multirow{2}{*}{\textbf{0.21}} & \multirow{2}{*}{No} \\
&calibrated		                 & 0.61 & 0.63 & 0.62 &  & \\
        \addlinespace[5pt]         
 
FD & \multicolumn{1}{c}{--}		                         & 0.00 & 1.00 & 0.00 & 0.29 & No\\
        \addlinespace[5pt]         

\multirow{2}{*}{MOG} & default                      & 0.21 & 0.90 & 0.35 & \multirow{2}{*}{5.17} & \multirow{2}{*}{\textbf{Yes}}\\
&calibrated                                      & 0.46 & 0.81 & 0.58 &  & \\
        \addlinespace[5pt]         

\multirow{2}{*}{FD+MOG} & default             & 0.54 & 0.68 & 0.60 & \multirow{2}{*}{5.33} & \multirow{2}{*}{No}\\
&calibrated                                    & 0.66 & 0.67 & 0.67 &  & \\
        \addlinespace[5pt]         

\multirow{2}{*}{MOG2} & default                & 0.10 & 0.96 & 0.17 & \multirow{2}{*}{2.31} & \multirow{2}{*}{\textbf{Yes}}\\
&calibrated                                     & 0.17 & 0.87 & 0.29 &  & \\
        \addlinespace[5pt]         

\multirow{2}{*}{FD+MOG2} & default            & 0.21 & 0.79 & 0.33 & \multirow{2}{*}{3.32} & \multirow{2}{*}{No}\\
&calibrated                                   & 0.22 & 0.93 & 0.35 &  & \\
        \addlinespace[5pt]         

\multirow{2}{*}{KNN} & default                   & 0.06 & 0.97 & 0.10 & \multirow{2}{*}{2.59} & \multirow{2}{*}{\textbf{Yes}}\\
&calibrated                                      & 0.08 & 0.96 & 0.14 & & \\
        \addlinespace[5pt]         

\multirow{2}{*}{FD+KNN} & default                             & 0.12 & 0.81 & 0.21 & \multirow{2}{*}{5.19} & \multirow{2}{*}{No}\\
&calibrated                                    & 0.13 & 0.95 & 0.22 & & \\
        \addlinespace[5pt]         

\multirow{2}{*}{GMG} & default                               & 0.17 & 0.89 & 0.29 & \multirow{2}{*}{8.44} & \multirow{2}{*}{No}\\
&calibrated                                      & 0.17 & 0.89 & 0.29 &  & \\
        \addlinespace[5pt]         

\multirow{2}{*}{FD+GMG} & default                             & 0.23 & 0.80 & 0.36 & \multirow{2}{*}{9.39} & \multirow{2}{*}{No}\\
&calibrated                                    & 0.24 & 0.95 & 0.38 &  & \\
        \addlinespace[5pt]         

\multirow{2}{*}{CNT} & default                               & 0.02 & 0.94 & 0.04 & \multirow{2}{*}{0.74} & \multirow{2}{*}{\textbf{Yes}}\\
&calibrated                                      & 0.42 & 0.81 & 0.56 &  & \\
        \addlinespace[5pt]         

\multirow{2}{*}{FD+CNT} & default                             & 0.02 & 0.73 & 0.03 & \multirow{2}{*}{1.67} & No\\
&calibrated                                    & 0.67 & 0.73 & 0.70 &  & \\
        \addlinespace[5pt]         

\multirow{2}{*}{FD+Thresh($\lambda$)} & default    & 0.85 & 0.93 & \textbf{0.88} & \multirow{2}{*}{0.47} & No\\
&calibrated                              & 0.95 & 0.90 & \textbf{0.93} &  & \\
        \addlinespace[5pt]         

\multirow{2}{*}{SubMax($\delta$)} & default                    & 0.00 & 0.07 & 0.00 & \multirow{2}{*}{0.96} & No\\
&calibrated                           & 0.00 & 0.65 & 0.01 &  & \\
        \addlinespace[5pt]         

\multirow{2}{*}{GSOC} & default                              & 0.07 & 1.00 & 0.13 & \multirow{2}{*}{267.37} & \multirow{2}{*}{\textbf{Yes}}\\
&calibrated                                     & 0.66 & 0.55 & 0.60 & & \\
        \addlinespace[5pt]         

\multirow{2}{*}{FD+GSOC} & default          & 0.26 & 0.87 & 0.41 & \multirow{2}{*}{296.03} & \multirow{2}{*}{No}\\
&calibrated                                   & 0.91 & 0.78 & 0.84 &  & \\

        \addlinespace[5pt]         

\multirow{2}{*}{LSBP} & default                              & 0.16 & 0.73 & 0.26 & \multirow{2}{*}{16.83} & \multirow{2}{*}{No}\\
&calibrated                                     & 0.18 & 0.68 & 0.29 &  & \\
        \addlinespace[5pt]         

\multirow{2}{*}{FD+LSBP} & default                            & 0.01 & 0.59 & 0.01 & \multirow{2}{*}{16.00} & \multirow{2}{*}{No}\\
&calibrated                                   & 0.01 & 0.56 & 0.01 &  & \\

        \bottomrule
    \end{tabular}
    \label{table:all_results1}
\end{table}
\begin{table}
    \centering
    \caption{Performance summary of all algorithms (Part 2).}
    \captionsetup{justification=centering}
    \begin{tabular}{llccccc}
        \toprule
        \multicolumn{2}{c}{\multirow{2}{*}[-1em]{Algorithms}} & \multicolumn{3}{c}{Segmentation accuracy} & Computation time & {Spatter robust?} \\
        \addlinespace[10pt]
        \cmidrule(lr){3-5} \cmidrule(lr){6-6} \cmidrule(lr){7-7}\\
        {} & {} & Precision & Recall & F\textsubscript{1}-score  & [\si{ms}] \\
        \midrule

\multirow{2}{*}{AdaptMean} & default                      & 0.00 & 0.85 & 0.00 & \multirow{2}{*}{1.21} & \multirow{2}{*}{No}\\
&calibrated                             & 0.00 & 1.00 & 0.00 &  & \\
        \addlinespace[5pt]         

\multirow{2}{*}{AdaptGauss} & default                  & 0.00 & 0.77 & 0.00 & \multirow{2}{*}{15.77} & \multirow{2}{*}{No}\\
&calibrated                         & 0.00 & 1.00 & 0.00 &  & \\
        \addlinespace[5pt]         

\multirow{2}{*}{FD+AdaptMean} & default  & 0.00 & 0.71 & 0.00 & \multirow{2}{*}{1.40} & \multirow{2}{*}{No}\\
&calibrated			         & 0.00 & 1.00 & 0.00 &  & \\
        \addlinespace[5pt]         

\multirow{2}{*}{FD+AdaptGauss} & default & 0.00 & 0.60 & 0.00 & \multirow{2}{*}{1.66} & \multirow{2}{*}{No}\\
&calibrated				 & 0.00 & 1.00 & 0.00 &  & \\
        \addlinespace[5pt]         

Otsu                       & \multicolumn{1}{c}{--}              & 0.00 & 0.96 & 0.01 & 0.72 & No\\
        \addlinespace[5pt]         

Li                         & \multicolumn{1}{c}{--}              & 0.00 & 1.00 & 0.00 & 9.21 & No\\
        \addlinespace[5pt]         

\multirow{2}{*}{Sauvola} & default                           & 0.00 & 0.91 & 0.00 & \multirow{2}{*}{8.36} & \multirow{2}{*}{No}\\
&calibrated                                  & 0.00 & 1.00 & 0.00 &  & \\
        \addlinespace[5pt]         

\multirow{2}{*}{FD+Sauvola} & default                         & 0.35 & 0.79 & 0.48 & \multirow{2}{*}{23.74} & \multirow{2}{*}{No}\\
&calibrated                                & 0.39 & 0.90 & 0.54 &  & \\
        \addlinespace[5pt]         

Triangle                   & \multicolumn{1}{c}{--}              & 0.01 & 1.00 & 0.02 & 0.69 & No\\
        \addlinespace[5pt]         

Yen & \multicolumn{1}{c}{--}  		 & 0.01 & 0.91 & 0.02 & 1.10 & No\\
        \addlinespace[5pt]         

isodata & \multicolumn{1}{c}{--}         & 0.00 & 1.00 & 0.00 & 1.12 & No\\
        \addlinespace[5pt]         

FD+Otsu     & \multicolumn{1}{c}{--}              & 0.62 & 0.59 & 0.60 & 1.15 & No\\
        \addlinespace[5pt]         

FD+Li       & \multicolumn{1}{c}{--}              & 0.36 & 0.94 & 0.52 & 7.02 & No\\
        \addlinespace[5pt]         

FD+Triangle & \multicolumn{1}{c}{--}              & 0.44 & 0.95 & 0.60 & 1.09 & No\\
        \addlinespace[5pt]         

FD+Yen      & \multicolumn{1}{c}{--}              & 0.73 & 0.84 & 0.78 & 1.51 & No\\
        \addlinespace[5pt]         

FD+isodata  & \multicolumn{1}{c}{--}              & 0.62 & 0.59 & 0.61 & 1.50 & No\\

        \bottomrule
    \end{tabular}
    \label{table:all_results2}
\end{table}

\end{document}